\documentclass[aps,twocolumn,showpacs,amsmath,amssymb,superscriptaddress,longbibliography,prb]{revtex4-2}
\usepackage[colorlinks=true,citecolor=blue,linkcolor=blue,hypertexnames=false]{hyperref}
\usepackage{amsmath}
\usepackage{dsfont}
\usepackage{hyperref}
\usepackage{textcomp}
\usepackage{tabularx}
\usepackage{graphicx,tikz}
\usepackage{amsfonts}

\newcommand{\cE}{\mathcal{E}}
\newcommand{\bk}{{\bf k}}
\newcommand{\bx}{{\bf x}}
\newcommand{\bq}{{\bf q}}

\newcommand{\bn}{{\bf n}}
\newcommand{\bd}{{\bf d}}

\newcommand{\bv}{{\bf v}}
\newcommand{\bP}{{\bf P}}

\newcommand{\be}{{\bf e}}

\newcommand{\n}{{\bf e}}

\newcommand{\bX}{{\bf X}}

\newcommand{\bdir}{{\bf e}}

\newcommand{\bsigma}{{\boldsymbol{\sigma}}}

\newcommand{\aCord}{a}
\newcommand{\bCord}{b}
\newcommand{\cCord}{c}

\def \ee{\end{equation}}
\def \bea{\begin{eqnarray}}
\def \eea{\end{eqnarray}}

\def \bea{\begin{eqnarray}}
\def \eea{\end{eqnarray}}

\begin{document}
\title{
    Gauge-invariant projector calculus for quantum state geometry \\ and applications to observables in crystals
}

\author{Johannes Mitscherling}
\thanks{These authors contributed equally.} 
\affiliation{Max Planck Institute for the Physics of Complex Systems, N\"othnitzer Str. 38, 01187 Dresden, Germany}
\affiliation{Department of Physics, University of California, Berkeley, CA 94720, USA}

\author{Alexander Avdoshkin}
\thanks{These authors contributed equally.}
\affiliation{Department of Physics, Massachusetts Institute of Technology, Cambridge, MA 02139, USA}
\affiliation{Department of Physics, University of California, Berkeley, CA 94720, USA}

\author{Joel E. Moore}
\affiliation{Department of Physics, University of California, Berkeley, CA 94720, USA}
\affiliation{Materials Sciences Division, Lawrence Berkeley National Laboratory, Berkeley, CA 94720, USA}
\date{\today}

\begin{abstract}
    The importance of simple geometrical invariants, such as the Berry curvature and quantum metric, constructed from the Bloch states of a crystal has become well-established over four decades of research. More complex aspects of geometry emerge in properties linking multiple bands, such as optical responses. In the companion work [arXiv:2409.16358], we identified novel multi-state geometrical invariants using an explicitly gauge-invariant formalism based on projection operators, which we used to clarify the relation between the shift current and the theory of electronic polarization among other advancements for second-order non-linear optics. Here, we provide considerably more detail on the projector formalism and the geometrical invariants arising in the vicinity of a specific value of crystal momentum. We combine the introduction to multi-state quantum geometry with broadly relevant algebraic relationships and detailed example calculations, enabling extensions toward future applications to topological and geometrical properties of insulators and metals.
    
\end{abstract}


\maketitle

\section{Introduction}

The integer quantum Hall effect and topological insulators can be understood as arising from topological invariants of Bloch states in crystals. The work of Thouless and collaborators to understand the Hall effect~\cite{Thouless1982} led eventually to an appreciation that the same geometrical objects that appear as integrands in topological invariants can also have physical implications of their own, including at a single point in momentum space. For example, the Berry curvature that integrates to the Chern number also determines, at each momentum point, the anomalous velocity in the semiclassical equations of motion for a Bloch wavepacket~\cite{Chang1995, Chang1996, Sundaram1999}. A practical advantage in expressing the Berry curvature, particularly when its derivatives need to be evaluated, or it needs to be integrated globally, came from the realization that it can be expressed in a gauge-invariant form using projection operators \cite{Avron1983}.

The two most studied geometrical invariants of this type are the quantum metric and Berry curvature, which can be understood as the real and imaginary parts of a single quantum geometric tensor \cite{Provost1980}. The present work is motivated by the realization that, even at a single point in momentum space, new geometrical objects emerge considering physical properties linking multiple bands, and that this {\it multi-state geometry} can be expressed and computed efficiently using projection operators. The physical consequences of that observation for shift currents and polarization, including calculations for some materials of current interest, are the focus of a short companion article; see Ref.~\onlinecite{prl2024}. In particular, we systematically developed the theory of these geometrical objects and explained how they characterize aspects of the Bloch states that go beyond the quantum metric and Berry curvature. In the following, we provide a detailed introduction to local single- and multi-state geometric invariants in combination with valuable identities and step-by-step examples, enabling the application of geometric concepts within the projector formalism to further observables in crystals.

As in classic works on the Berry curvature \cite{Avron1983} and successfully applied in a broad class of applications \cite{Pozo2020, Graf2021, Mera2021, Mera2022, Avdoshkin2022, Avdoshkin2024, Antebi2024}, the basic idea of using projectors built from Bloch states instead of the Bloch states themselves is to avoid the ambiguous momentum-dependent phase. Avoiding this gauge dependence not only simplifies standard perturbation-theoretical approaches using Green's function methods \cite{Dupuis2023} and allows for their efficient computation, as we will discuss in the context of non-linear optical responses, but allows for a geometrical interpretation of the complex projective bundle that the Bloch states form over the Brillouin zone. The core of this approach is using differential geometry to establish relationships among projector objects, identifying a small set of independent geometric invariants and inequalities that underpin observable relationships. In many circumstances, observables can be expressed as scalar quantities (e.g., trace or determinant) derived from projectors, which suggest a geometrical form that illuminates physical content beyond topological invariants. Conversely, any such scalar is, in principle, observable, providing pathways toward unexplored observables and their physical phenomenology. This work focuses on a universal geometric approach to quantum states based on projectors arising from a Bloch Hamiltonian of a non-interacting multiorbital lattice system. Here, orthogonal, possibly degenerate, projectors parameterized by quasimomentum complement the information yielded by the energy bands. We explicitly demonstrate that novel geometric invariants arise in several examples, which capture quantum state properties beyond the quantum geometric tensor and objects that reduce to it, such as the quantum metric dipole, Berry curvature dipole, and Riemannian tensors. Identifying and physically interpreting geometric invariants is a subject of ongoing work \cite{Topp2019, Topp2021, Tai2023, Mera2022, Chen2022, Kashihara2023, Verma2024, Ahn2020, Ahn2022, Bouhon2023, Avdoshkin2022, Avdoshkin2024, Antebi2024, Holder2020, Kaplan2023, Wojciech2024, Kruchkov2023, onishi2024quantum, onishi2024universal, onishi2024fundamental, Bradlyn2024}. 
    
The paper is structured as follows. In Sec.~\ref{sec:projectors}, we introduce the basic concepts and quantities for quantum state geometry and projector calculus. After presenting the recently emerging terminology for quantum state geometry beyond the quantum geometric tensor in Sec.~\ref{sec:ProjIntro} to \ref{sec:MinimalStructures}, we summarize the main geometric invariants that are currently intensively studied in Sec.~\ref{sec:ImportantGeometricInvariants}; see an overview of these quantities in projector form in Tab.~\ref{tab:GeometricInvariants}. We close by various valuable projector identities for applying the projector formalism in the derivations of observables in Sec.~\ref{sec:UsefulProjectorIdentities}. We demonstrate the formalism in two applications in Sec.~\ref{sec:Applications}, demonstrating the close connection to the modern theory of polarization in Sec.~\ref{sec:Polarization} and exemplifying the derivation of a non-linear response function via a detailed derivation of the injection and shift current in Sec.~\ref{sec:ShiftCurrent}. We provide the main conclusion and a short outlook in Sec.~\ref{sec:Conclusion}. Further technical details are given in the appendix.

\begin{table*}[t!]
    \centering
    \begin{tabular}{l l l 
    }
         & {\bf Single-state} & {\bf Two-state} 
         \\[1mm]
         \hline \\[-2mm]
       Quantum geometric tensor  &  $Q_{\alpha\beta}=\text{tr}\big[\hat P \,\partial_\alpha \hat P\, \partial_\beta P\big]$ & $Q^{mn}_{\alpha\beta} = \text{tr}\big[\hat P_n \,\partial_\alpha \hat P_m \,\partial_\beta \hat P_n\big]$
       \\[1mm]
       Quantum geometric connection & $Q_{\alpha;\beta\gamma} = \text{tr}\big[\hat P\,\partial_\alpha \hat P\,\partial_\beta\partial_\gamma\hat P\big]$& $C^{mn}_{\alpha;\beta\gamma} =\text{tr}\big[\hat P_n\,\partial_\beta\hat P_{m}\big(\partial_\alpha\partial_\gamma\hat P_n+\partial_\alpha\hat P_m\,\partial_\gamma\hat P_n\big)\big]$
       \\[1mm] \hline\\[-2mm]
       Quantum metric & $g_{\alpha\beta}=\frac{1}{2}\text{tr}\big[\partial_\alpha \hat P\,\partial_\beta \hat P\big]$& $g^{mn}_{\alpha\beta} = \frac{1}{2}\text{tr}\big[\partial_\alpha \hat P_m\,\partial_\beta \hat P_n\big]$
       \\[2mm]
       Berry curvature & $\Omega_{\alpha\beta} = i \,\text{tr}\big[\hat P\,\partial_\alpha\hat P\,\partial_\beta\hat P\big]-(\alpha\leftrightarrow \beta)$ & $\Omega^{mn}_{\alpha\beta} = i\,\text{tr}\big[\hat P_n\,\partial_\alpha \hat P_m\,\partial_\beta \hat P_n\big]-(\alpha\leftrightarrow \beta)$
       \\[2mm]
       Quantum metric dipole & $\partial_\alpha g_{\beta\gamma} = \frac{1}{2}\text{tr}\big[\partial_\beta\hat P\,\partial_\alpha\partial_\gamma\hat P\big]+(\beta\leftrightarrow\gamma)$& $\partial_\alpha g^{mn}_{\beta\gamma}=\frac{1}{2}\text{tr}\big[\partial_\alpha(\partial_\beta \hat P_m\,\partial_\gamma \hat P_n)\big]$
       \\[2mm]
       Berry curvature dipole & $\partial_\alpha \Omega_{\beta\gamma} = i\,\text{tr}\big[(\hat P\,\partial_\beta\hat P-\partial_\beta\hat P\,\hat P)\partial_\alpha\partial_\gamma \hat P\big]$& $\partial_\alpha \Omega^{mn}_{\beta\gamma}= i\,\text{tr}\big[\partial_\alpha(\hat P_n\,\partial_\beta \hat P_m\,\partial_\gamma \hat P_n)\big]-(\beta\leftrightarrow \gamma)$
       \\[1mm] & $\hspace{15mm}-(\beta\leftrightarrow\gamma)$& 
       \\[1mm] \hline\\[-2mm]
       Skewness tensor & $\text{Im}\,Q_{(\alpha;\beta\gamma)}=\frac{1}{3}\text{Im}\big[Q_{\alpha;\beta\gamma}+Q_{\beta;\gamma\alpha}+Q_{\gamma;\alpha\beta}\big]$& ---
       \\[1mm]
       Torsion tensor & --- & $T^{mn}_{\beta;\alpha\gamma} = \text{tr}\big[\hat P_n\,\partial_\beta \hat P_m\,\partial_\alpha \hat P_m\,\partial_\gamma \hat P_n\big]-(\alpha\leftrightarrow \gamma)$ \\[2mm]
       Shift vector & --- & $R^{mn}_{\alpha\beta}(\bk) = i\,C^{mn}_{\alpha;\beta\beta}(\bk)/Q^{mn}_{\beta\beta}(\bk)$
    \end{tabular}
    \caption{We present all geometric invariants introduced and discussed throughout this paper in their projector form.}
    \label{tab:GeometricInvariants}
\end{table*}


\section{Projector calculus for quantum state geometry} \label{sec:projectors}

We introduce the essential techniques of projector calculus for quantum state geometry, focusing on the aspects that enable a profound understanding of the applications on the polarization distribution and shift current presented in the short companion article \cite{prl2024} and guide future applications to other observables in crystals. For simplicity and concreteness, we restrict ourselves to non-interacting Bloch Hamiltonians and leave a comprehensive description of quantum state geometry in the context of interacting multi-orbital lattice systems for future work. For a first introduction in this direction, we refer to Ref.~\onlinecite{Antebi2024}, which revealed novel geometric invariants, such as the Berry curvature variance, in the context of the Drude weight of interacting flatband systems. 

In the following, we introduce the projectors for non-degenerate and degenerate bands, enabling us to express scalar observables in the context of quantum state geometry. We report on recent progress in identifying the minimal quantum geometric structures, and we provide a summary of geometric invariants of current interest; see Tab.~\ref{tab:GeometricInvariants}. We conclude this section with useful projector identities that simplify the subsequent applications.

\subsection{Projectors for non-degenerate bands}
\label{sec:ProjIntro}

We start with a Bloch Hamiltonian $\hat H(\bk)$ involving $N$ bands as a function of lattice momentum $\bk$ in d dimensions. When diagonalized for a fixed momentum, we obtain the eigenvalues $E_n(\bk)$ and corresponding orthonormal eigenvectors $|u_n(\bk)\rangle$ with band index $n$. In the case of only non-degenerate bands, we construct the band projectors $\hat P_n(\bk)$ as the tensor product between the eigenvectors $|u_n(\bk)\rangle$ and its conjugate transpose $\langle u_n(\bk)|$, resulting in 
\begin{align}
    \hat P_n(\bk) = |u_n(\bk)\rangle\langle u_n(\bk)| \, .
\end{align}
No specific $U(1)$ gauge choice is required for $|u_n(\bk)\rangle$ within this construction as long as $\langle u_n(\bk)|$ is obtained from the corresponding $|u_n(\bk)\rangle$ since 
\begin{align}
    \hat P_n(\bk)\rightarrow e^{i\phi_n(\bk)}\,\hat P_n(\bk)\,e^{-i\phi_n(\bk)}= \hat P_n(\bk) \, .
\end{align}
The gauge-invariant projector calculus harnesses the following four main properties. The projectors are hermitian, idempotent, and orthogonal; that is, they satisfy 
\begin{align}
    \label{eqn:basics1}
    \big(\hat P_n(\bk)\big)^\dagger &= \hat P_n(\bk) 
\end{align}
and
\begin{align}
    \label{eqn:basics2}
     \hat P_n(\bk) \,\hat P_m(\bk) &= \delta_{nm}\,\hat P_m(\bk) \, ,
\end{align}
where the product of two projectors is defined as ordinary matrix multiplication and $\delta_{nm}$ is the Kronecker delta. As the projectors are defined with respect to the eigenstates of a given Bloch Hamiltonian $\hat H(\bk)$, they further satisfy
\begin{align}
    \label{eqn:basics3}
    \hat H(\bk)\, \hat P_n(\bk) &= E_n(\bk)\, \hat P_n(\bk)
\end{align}
and
\begin{align}
    \label{eqn:basics4}
    \sum_n \hat P_n(\bk) &= \mathds{1}_N \, ,
\end{align}
under summation over all $N$ bands. Traces over products of projectors are understood as traces of the respective matrix. For instance, the Bloch-state overlap at different momenta and bands in projector form reads
\begin{align}
    \text{tr}\big[\hat P_n(\bk) \hat P_m(\bk')] = |\langle u_n(\bk)|u_m(\bk')\rangle|^2 \, .
\end{align}
A key advantage in working with projectors instead of Bloch wave functions is the possibility of an efficient analytical and numerical evaluation of {\it geometric invariants}, that is, combinations of multiple projectors and their derivatives at the same or different momenta. The main reason for this efficiency is the elimination of gauge ambiguity of the band basis. In particular, derivatives of projectors, formally defined as 
\begin{align}
    \partial_\alpha \hat P_n(\bk) \equiv \lim_{\delta\rightarrow 0} \frac{1}{\delta}\big[\hat P_n(\bk+\delta\, \be_\alpha)-\hat P_n(\bk)\big] \, ,
\end{align}
with the short notation of the derivative $\partial_\alpha\equiv \partial/\partial k_\alpha$ in direction $\be_\alpha$, are well-defined and straightforwardly numerically implemented, see Appendix~\ref{sec:numericalEvaluation}. 

Closed analytic forms of the projector in terms of the Bloch Hamiltonian without the need of any diagonalization are particularly useful. The projectors onto the lower and upper band of a two band system expressed in the form $\hat H(\bk) = d_0(\bk) + \bd(\bk)\cdot \bsigma$ with Pauli matrices $\bsigma = (\sigma_x,\sigma_y,\sigma_z)$ serve as an important example and read 
\begin{align}
    \hat P_\pm(\bk) = \frac{1}{2}\Big[\mathds{1}_2\pm \frac{\bd(\bk)}{|\bd(\bk)|}\cdot \bsigma\Big] \, .
\end{align}
Recently, a growing number of results are available that offer closed analytic formulas for projectors directly obtained in terms of the Hamiltonian for more than two bands without any explicit diagonalization of the Hamiltonian \cite{Pozo2020, Graf2021}, which we discuss in more detail in Appendix~\ref{sec:Pozo2020}. Expressing geometric invariants via derivatives of the Bloch Hamiltonian itself is possible for special cases, such as the Berry curvature \cite{Bernevig2013}, via the identities introduced in Sec.~\ref{sec:UsefulProjectorIdentities}.

\subsection{Projectors for degenerate bands and larger subsets}

The projector formalism naturally includes the case of degenerate bands and, thus, offers their theoretical description on an equal footing. If the band is formed by $|u_{n s}(\mathbf{k})\rangle$ with $s = 1,\,...\,,M$ indexing the states within the $M$-fold degenerate band (we assume degeneracy at all $\mathbf{k}$), the projector on the degenerate subspace reads
\begin{align}
    \hat P_n(\mathbf{k}) = \sum_{s=1}^M | u_{ns}(\mathbf{k}) \rangle \langle u_{ns}(\mathbf{k}) | \, .
\end{align}
The same basic properties \eqref{eqn:basics1} to \eqref{eqn:basics4} hold. These projectors explicitly respect the required $U(M)$ gauge invariance of the degenerate band. In addition, it might be convenient to combine multiple bands $n_1$ to $n_M$ into one subspace of quantum states of interest, 
\begin{align}
    \hat P_{(n_1...n_M)}(\bk) = \sum_{i = 1}^M \hat P_{n_i}(\bk) \, ,
\end{align}
which still satisfy the projector properties \eqref{eqn:basics1} and \eqref{eqn:basics2}. The most common example is the projector onto occupied states, which takes the form $\hat P_\text{occ}(\bk)=\sum_{n\in \text{occ}}\hat P_n(\bk)$. 

\subsection{Global and local geometric invariants}

Being gauge invariant and hermitian, projectors are, in principle, measurable and, thus, offer minimal building blocks to construct observables. The projectors $\hat P(\bk)$ constructed from a given Bloch Hamiltonian $\hat H(\bk)$ capture the nontrivial momentum-dependence of the eigenstates arising from the complex interplay of the {\it orbitals}, including atomic orbitals, spin, sites within the unit cells, and other potential quantum numbers relevant in the system under investigation. 

A natural way to form scalar observables is to take traces of products of projectors, e.g.,
\begin{align} \label{eq:global_p_expr}
    \text{tr}\big[\hat P_1(\bk_1) \,\hat P_2(\bk_2) \,\hat P_1(\bk_3) \,\hat P_3(\bk_4)\big]\, ,
\end{align}
involving multiple projectors of the same or different subsets and momenta. In the two cases we consider as applications in Sec.~\ref{sec:Polarization} and \ref{sec:ShiftCurrent}, e.g., the polarization distribution and optical responses, the final expression involve projector combinations at only a single momentum $\bk$ in combination with derivatives of the projectors, e.g.,
\begin{align} \label{eq:local_p_expr}
    \text{tr}\Big[ \hat P^{}_1(\mathbf{k}) \,\big( \partial^{}_{a}\hat  P^{}_2(\mathbf{k}) \big)\, \big(  \partial^{}_b \hat P^{}_1(\mathbf{k}) \big) \,\big( \partial^{}_{c_1}\partial^{}_{c_2}\partial^{}_{c_3}\hat P^{}_3(\mathbf{k})\big) \Big] \,.
\end{align}
Expressions like Eq.~\eqref{eq:global_p_expr} involving projectors at arbitrarily separated points on the Brillouin zone are called \emph{global geometric invariants}. Expressions like Eq.~\eqref{eq:local_p_expr} involving derivatives of projectors, that is, only infinitesimally separated in $\bk$, are called \emph{local geometric invariants}. Both should yield a complete characterization of the Bloch states with some caveats in the case of local invariants \cite{Avdoshkin2022}. The goal of {\it quantum state geometry} is to identify all independent structures and their relationships. One of the main objectives is to identify the minimal number of structures that can appear in a specified physical context, such as optical responses and polarization distribution.

\subsection{Intrinsic and extrinsic geometry}

When determining the minimal set of independent geometric structures, it is essential to identify relations between different geometric objects. We would like to point out that these relations depend on whether the whole manifold of possible quantum states for a given number of bands is considered or just its submanifold made up by the state corresponding to the Bloch states, that is, those states in the whole manifold that are in the image of the projector $\hat P(\bk)$. This distinction becomes relevant when the (real) dimension of the complex projective space formed by all possible states is larger than the dimension of the Brillouin zone, i.e., $2(N-1)>d$ with $N$ bands in $d$ dimensions for a non-degenerate band.

This aspect is well established in Euclidean geometry, e.g., for surfaces in $\mathds{R}^3$. At every point, the surface is described by two principal curvatures. Since they do not change under isometries, they are individual geometric invariants of the surface embedding into the 3-dimensional space. However, only the product of the curvatures is intrinsic to the surface, i.e., it can be reconstructed from the induced metric of $\mathds{R}^3$ onto the surface. Thus, embedding is crucial when characterizing the geometry, which can give rise to further geometric invariants.

We refer to the geometric invariants on the ambient space---the manifold of all possible quantum states---as {\it intrinsic} and all other geometric information, which characterizes the embedding as {\it extrinsic}. In Ref.~\cite{Avdoshkin2022}, it was shown for a single non-degenerate projector that the only intrinsic object is the quantum geometric tensor and objects of the form with second- and higher-order derivatives completely characterize the remaining extrinsic geometry; see Eqs.~\eqref{eqn:Qalpha_ab} and \eqref{eqn:localGeoInv}. The intrinsic objects for degenerate states are described in Sec. \ref{sec:non-abelian}. The geometry of multi-state objects is not yet sufficiently understood for separation into intrinsic and extrinsic invariants. Progress can be made by considering a Taylor expansion of global geometric invariants to identify those that fully describe the embedding.  

\subsection{Minimal quantum geometric structures}
\label{sec:MinimalStructures}

If only a single non-degenerate band $\hat P_m$ with band index $m$ is involved, it was shown by one of us~\cite{Avdoshkin2022} that all global geometric invariants can be reduced to three-point functions of the form 
\begin{align} \label{eq:global_inv}
    \text{tr}\big[\hat P^{}_m(\mathbf{k}_1)\, \hat P^{}_m(\mathbf{k}_2)\, \hat P^{}_m(\mathbf{k}_3)\big]\,,
\end{align}
involving band projectors at three distinct momenta. Furthermore, all local geometric invariants reduce to a series of objects of the form 
\begin{align}
    \label{eqn:localGeoInv}
    Q_{\alpha;\beta_1 ... \beta_n}\!(\mathbf{k}) \equiv \text{tr}\Big[\hat P(\mathbf{k})\big(\partial_\alpha \hat P(\mathbf{k}) \big)\big(\partial^{}_{\beta_1}...\,\partial^{}_{\beta_n}\!\hat P(\mathbf{k})\big)\!\Big]  ,
\end{align}
involving the band projector and its first-order and $n$-th derivative. Note that by construction $Q_{\alpha;\beta_1 ... \beta_n}$ is symmetric in the $b_i$ indices. In contrast, similar statements for degenerate bands and multi-state quantum geometric structures involving projectors of multiple bands are more elaborate. A recent analysis by one of us showed that three-point functions are still sufficient for a single $M$-times degenerate band but show a much richer internal structure~\cite{Avdoshkin2024}. The part of the relevant structure discussed in the joint submission~\cite{prl2024} and throughout this paper comes from Slater determinants and is equivalent to the non-degenerate case via Pl\"ucker embedding; see Sec.~\ref{sec:Polarization}. A complete understanding of the quantum geometric structure of multi-state (multiband) observables is still missing but is highly relevant for applications. For example, resonant optical responses involve two bands $\hat P_n$ and $\hat P_m$ when their band dispersions are on resonance, i.e., when $E_n(k)-E_m(k) = \omega$. New geometric objects have been identified and studied~\cite{Ahn2022, Wojciech2024, romero2024n} in line with the discussion presented in the joint submission~\cite{prl2024}. Despite recent progress, a comprehensive description is still lacking. 

\subsection{Important geometric invariants}
\label{sec:ImportantGeometricInvariants}

We summarize several geometric invariants that have recently attracted attention and write them in projector form. We provide a summary in Tab.~\ref{tab:GeometricInvariants}. We start with the quantum geometric tensor, involving the quantum metric and Berry curvature, and its multi-state variant. We continue with the quantum geometric connection, relevant for the third cumulant, or skewness, of the polarization distribution and the shift current, as introduced in the joint submission~\cite{prl2024}, and the quantum metric and Berry curvature dipoles. We conclude by stressing the relation to the torsion tensor, a novel geometric quantity in systems with at least three bands \cite{Ahn2022, Wojciech2024}. Since we only focus on local geometric invariants, we introduce the short notation $\hat P \equiv \hat P(\bk)$, omitting the momentum dependence of the projector.

\subsubsection{Quantum geometric tensor}

The {\it (single-state) quantum geometric tensor} in projector form reads
\begin{align}
    \label{eqn:Qalpha_ab}
    Q_{\alpha\beta} \equiv \text{tr}\big[\hat P^{}\,(\partial^{}_\alpha\hat P^{})\,(\partial^{}_\beta\hat P^{})\big] \, ,
\end{align}
involving a projector onto a (non-)degenerate band, occupied states, or other sets of quantum states. Besides the quantum geometric tensor expressed in the Bloch states for a non-degenerate band, see Appendix~\ref{sec:GeometryNonDeg}, the projector form \eqref{eqn:Qalpha_ab} has been extensively used in recent years, for instance, in the context of fractional quantum Hall physics; see, e.g., \cite{Roy2014, Mera2021, Ledwith2023}. Decomposing the quantum geometric tensor into its symmetric and antisymmetric part with respect to the tensor indices $\alpha$ and $\beta$ or, equivalently, into its real and imaginary part, we identify the quantum metric $g_{\alpha\beta}$ and Berry curvature $\Omega_{\alpha\beta}$ associated with the projector $\hat P^{}$ as
\begin{align}
    \label{eqn:decompositionQab}
    Q_{\alpha\beta}=g_{\alpha\beta}-\frac{i}{2}\Omega_{\alpha\beta}
\end{align}
with
\begin{align}
    \label{eqn:quantummetric}
    &g_{\alpha\beta}(\bk) 
    \equiv \frac{1}{2}\text{tr}\Big[\big(\partial^{}_\alpha\hat P\big)\big(\partial^{}_\beta\hat P\big)\Big] \, , \\ 
    \label{eqn:Berrycurvature}
    &\Omega_{\alpha\beta}(\bk) 
    \equiv i\,\text{tr}\Big[\hat P\big(\partial^{}_\alpha\hat P\big)\big(\partial^{}_\beta\hat P\big)-\hat P\big(\partial^{}_\beta\hat P\big)\big(\partial^{}_\alpha\hat P\big)\Big]\, .
\end{align}
We emphasize that the choice of quantum states of interest via the projector defines the quantum geometric tensor and, thus, the quantum metric and Berry curvature. The quantities for different projectors are generally not simply related; see, e.g., for the quantum metric of individual bands and the occupied states \cite{Peotta2015, Mera2022}. 

As the first multi-state geometric invariant, we introduce the {\it two-state quantum geometric tensor}, which naturally generalizes the expression \eqref{eqn:Qalpha_ab},
\begin{align}
    \label{eqn:Qnm}
    Q^{mn}_{\alpha\beta} &\equiv \text{tr}\Big[\hat P^{}_{n}\,\big(\partial^{}_\alpha\hat P^{}_{m}\big)\big(\partial^{}_\beta\hat P^{}_{n}\big)\Big] 
\end{align}
involving two (potentially) distinct projectors $\hat P_n$ and $\hat P_m$, usually referring to projectors on two different bands. For this, note the difference in the index of the middle projector. The decomposition of the two-state quantum geometric tensor in analogy to Eq.~\eqref{eqn:decompositionQab} into its symmetric and antisymmetric contribution naturally generalizes the quantum metric and Berry curvature to their two-state quantities, i.e., 
\begin{align}
    Q^{mn}_{\alpha\beta} = g^{mn}_{\alpha\beta}-\frac{i}{2}\Omega^{mn}_{\alpha\beta} \, 
\end{align}
with
\begin{align}
    \label{eqn:gmn_ab}
    &g^{mn}_{\alpha\beta} \equiv \frac{1}{2}\text{tr}\Big[\big(\partial_\alpha\hat P^{}_m(\bk)\big)\big(\partial_\beta \hat P^{}_n(\bk)\big)\Big] \, , \\
    \label{eqn:Omn_ab}
    &\Omega^{mn}_{\alpha\beta} \equiv i\,\text{tr}\Big[\hat P^{}_{n}\big(\partial^{}_\alpha\hat P^{}_{m}\big)\!\big(\partial^{}_\beta\hat P^{}_{n}\big)\!-\!\hat P^{}_{n}\big(\partial^{}_\beta\hat P^{}_{m}\big)\!\big(\partial^{}_\alpha\hat P^{}_{n}\big)\!\Big].
\end{align}
We note that the two-state quantum geometric tensor decomposes into the diagonal part given by the single-state quantum geometric tensor given in Eq.~\eqref{eqn:Qalpha_ab} and a purely two-state contribution, 
\begin{align}
    Q^{mn}_{\alpha\beta} = \delta_{nm}\,Q^{n}_{\alpha\beta}-\text{tr}\Big[\hat e_\alpha^{nm}\,\hat e_\beta^{mn}\Big] \, .
\end{align}
which we expressed in terms of 
\begin{align}
    \label{eqn:emna}
    \hat e^{mn}_\alpha=i\,\hat P_m\big(\partial_\alpha \hat P_n\big)\hat P_n \, .
\end{align}
Note that $\hat e^{nn}_\alpha = 0$ individually, which follows from projector identity \eqref{eqn:basics2}. The purely two-state contribution $\text{tr}[\hat e^{nm}_\alpha \hat e^{mn}_\beta]$ yields the product of non-Abelian Berry connections, see Appendix~\ref{sec:GeometryNonDeg}, and is, thus, directly connected to interband transitions in optical responses \cite{Ahn2022}. Using the completeness of the band projectors~\eqref{eqn:basics4}, the single- and two-state quantum geometric tensor are related via
\begin{align}
    \label{eqn:sum_Qmn}
    \sum_{m\neq n} Q^{mn}_{\alpha\beta} = - Q^n_{\alpha\beta} \, .
\end{align}
Note that the negative-sign convention of the two-state quantum geometric tensor for $n\neq m$ arises from the form given in Eq.~\eqref{eqn:Qnm} fixing the relation $Q^{nn}_{\alpha\beta}=Q^n_{\alpha\beta}$ via the commonly used definition in Eq.~\eqref{eqn:Qalpha_ab}. We conclude by noticing that $g^{mn}_{\alpha\beta} = g^{(mn)}_{(\alpha\beta)}$ and $\Omega^{mn}_{\alpha\beta} = \Omega^{[mn]}_{[\alpha\beta]}$ for $n\neq m$, where we denote the symmetrization and antisymmetrization of the indices as $(\alpha\beta)$ and $[\alpha\beta]$, respectively. Thus, the symmetry with respect to the momentum derivative imposes the symmetry with respect to the band indices, which offers straightforward simplifications in analytic calculations \cite{Mera2022, prl2024}.

\subsubsection{Quantum geometric connection}

We continue by considering the second local geometric invariant in Eq.~\eqref{eqn:localGeoInv}. For a given projector $\hat P$ we have the {\it (single-state) quantum geometric connection}
\begin{align}
    \label{eqn:Qalphabetagamma}
    Q_{\alpha;\beta\gamma} \equiv \text{tr}\Big[\hat P\,\big(\partial^{}_\alpha \hat P\big)\big(\partial^{}_\beta\partial^{}_\gamma\hat P\big)\Big] \, ,
\end{align}
which involves a first- and second-order derivative of the projector. Note that $Q_{\alpha;\beta\gamma}=Q_{\alpha;\gamma\beta}$. The quantum geometric connection is a complex-valued three-tensor and, therefore, is expected to yield a more complex structure than the quantum geometric tensor, including novel independent geometric information about the quantum states. As shown in Ref.~\cite{Avdoshkin2022} by one of us, and extended to degenerate bands in the joint submission \cite{prl2024}, the imaginary part of $Q_{\alpha;\beta\gamma}$ is related to the third cumulant of the polarization distribution $\langle \hat X_\alpha \hat X_\beta \hat X_\gamma\rangle_c$ under integration over the Brillouin zone (BZ); see details in Sec.~\ref{sec:Polarization}. We note that 
\begin{align}
    \label{eqn:intQ}
    \text{Im}\,Q_{\alpha;\beta\gamma} = \text{Im}\,Q_{(\alpha;\beta\gamma)} - \frac{1}{6}\big(\partial^{}_\beta\,\Omega_{ac}+\partial^{}_\gamma\,\Omega_{\alpha\beta}\big) \, ,
\end{align}
which follows from vanishing combinations of projector derivatives; see identities in Sec.~\ref{sec:vanishingProjectorCombinations}. Thus, we find $\int_\text{BZ} \text{Im}\, Q_{\alpha;\beta\gamma} = \int_\text{BZ} \text{Im}\, Q_{(\alpha;\beta\gamma)}$ since the boundary contribution involving the Berry curvature $\Omega_{\alpha\beta}$ vanishes. As Eq.~\eqref{eqn:intQ} suggests, the quantum geometric connection contains the {\it quantum metric dipole} and {\it Berry curvature dipole}. Indeed, we have
\begin{align}
    &\partial_\alpha\, g_{\beta\gamma} = \text{Re}\Big[Q_{\beta;\alpha\gamma}+Q_{\gamma;\alpha\beta}\Big]  \, ,\\ 
    &\partial_\alpha\, \Omega_{\beta\gamma} = -2\,\text{Im}\Big[Q_{\beta;\alpha\gamma}-Q_{\gamma;\alpha\beta}\Big] \, ,
\end{align}
under symmetrization with respect to the first and last index of the quantum geometric connection. The quantum metric and Berry curvature dipole have been related, e.g., to nonlinear transport in a topological antiferromagnet \cite{Wang2023} and the quantum nonlinear Hall Effect \cite{Sodemann2015}, respectively.

As for the quantum geometric tensor, we present the multi-state generalization of Eq.~\eqref{eqn:Qalphabetagamma} and define the {\it two-state quantum geometric connection}
\begin{align}
    \label{eqn:Cmn_abg}
    C^{mn}_{\alpha;\beta\gamma} 
    \!\equiv\!\text{tr}\bigg[\hat P^{}_{n}\!\big(\partial^{}_\beta\hat P^{}_{m}\big)\!\Big[\!\big(\partial^{}_\alpha\partial^{}_\gamma\hat P^{}_{n}\big)\!+\!\big(\partial^{}_\alpha\hat P^{}_{m}\big)\!\big(\partial^{}_\gamma\hat P^{}_{n}\big)\!
    \Big]\!\bigg], 
\end{align}
which involve a second-order derivative in combination with the product of three first-order derivatives of the projectors in a particular combination of indices $n$ and $m$, usually related to band indices. This second term vanishes in the single-state (band-diagonal) component, see Sec.~\ref{sec:vanishingProjectorCombinations}, so that $C^{nn}_{\alpha;\beta\gamma}=Q^n_{\alpha;\beta\gamma}$ given in Eq.~\eqref{eqn:Qalphabetagamma} with respect to the band projector $\hat P_n$. This identification fixes our sign convention. We express the purely two-state (interband) component in terms of $\hat e^{nm}_\alpha$ and its covariant derivative $\nabla^{}_\alpha\hat e^{mn}_\gamma\equiv \hat P^{}_m\big(\partial^{}_\alpha \hat e^{mn}_\gamma\big)\hat P^{}_n$ as introduced in Ref.~\onlinecite{Ahn2022} and find
\begin{align}
    C^{mn}_{\alpha;\beta\gamma}&=\delta_{nm}\,Q^n_{\beta;\alpha\gamma}-\text{tr}\big[\hat e_\beta^{nm}\,\nabla^{}_\alpha \hat e_\gamma^{mn}\big] \, .
\end{align}
This form highlights the connection to higher-order optical responses as presented by Ahn {\it et al.} \cite{Ahn2022}, where they interpreted transition dipole moment matrix elements as tangent vectors $\hat e^{mn}_\alpha$ with projector form given in Eq.~\eqref{eqn:emna}. 

We have a closer look at the tensor properties of the two-state quantum geometric connection. When decomposing $C^{mn}_{\alpha;\beta\gamma}$ for $n\neq m$ uniquely into its symmetric and antisymmetric components with respect to the band indices and two last derivative directions, we find that $C^{(mn)}_{\alpha;(\beta\gamma)}$ and $C^{[mn]}_{\alpha;[\beta\gamma]}$ are proportional to the dipole of the two-state quantum metric \eqref{eqn:gmn_ab} and Berry curvature \eqref{eqn:Omn_ab}, respectively. The remaining $C^{(mn)}_{\alpha;[\beta\gamma]}$ and
$C^{[mn]}_{\alpha;(\beta\gamma)}$ determine the shift current under linear and circular polarized illumination, respectively; see Ref.~\cite{prl2024} and Sec.~\ref{sec:ShiftCurrent} for a detailed discussion. The real and imaginary parts of $C^{mn}_{\alpha;\beta\gamma}$ are symmetric and antisymmetric in the band indices for $n\neq m$, respectively. We conclude by noticing that the summation over the off-diagonal band indices is not simply related to the single-state quantum geometric connection but yields a second term, i.e.,
\begin{align}
    \label{eqn:sum_CmnAppendix}
    \sum_{m\neq n} C^{mn}_{\alpha;\beta\gamma} = &- Q^n_{\beta;\alpha\gamma} \nonumber \\ &\!+ \text{tr}\bigg[\hat P_n\Big[\!\sum_{m\neq n}\!\! \big(\partial_\beta \hat P_m\big)\!\big(\partial_\alpha \hat P_m\big)\!\Big]\!\big(\partial_\gamma\hat P_n\big)\!\bigg] ,
\end{align}
in contrast to the two-state quantum geometric tensor in Eq.~\eqref{eqn:sum_Qmn}. Note that the second term arises only in systems with more than two bands and vanishes in a two-band system; see Sec.~\ref{sec:vanishingProjectorCombinations}.

Single-state 4-derivative objects have been considered in \cite{hetenyi2023fluctuations, onishi2024geometric}.

\subsubsection{Torsion tensor}

The two-state quantum geometric connection yields novel geometric information about the Bloch state manifold in systems with more than two bands, the torsion. Following our insights presented in the joint submission \cite{prl2024}, we introduce the {\it torsion tensor} in projector form as 
\begin{align}
    T^{mn}_{\beta;\alpha\gamma} \equiv \text{tr}\Big[\hat P_n \big(\partial_\beta \hat P_m\big)\big(\partial_\alpha\hat P_m\big)\big(\partial_\gamma\hat P_n\big)\Big]-(\alpha\leftrightarrow\gamma) \, ,
\end{align}
where we denote antisymmetrization in $\alpha$ and $\gamma$ as the second term. This form agrees with the original definition \cite{Ahn2022} in terms of the tangent vector $\hat e^{mn}_\alpha$ using Eq.~\eqref{eqn:emna}, 
\begin{align}
    T^{mn}_{\beta;\alpha\gamma} = \text{tr}\big[\hat e^{nm}_\beta\big(\nabla^{}_\alpha\,\hat e^{mn}_\gamma-\nabla^{}_\gamma\,\hat e^{mn}_\alpha-[\hat e^{mn}_\gamma,\hat e^{mn}_\alpha]\big)\big]
\end{align}
and is related to the antisymmetrization in the first and last index of the two-state quantum geometric connection, i.e.
\begin{align}
    T^{mn}_{\beta;\alpha\gamma} = C^{mn}_{\gamma;\beta\alpha}-C^{mn}_{\alpha;\beta\gamma} \, .
\end{align}
Note that the torsion tensor vanishes for $n=m$ as well as for two-band systems as follows from the projector identity~\eqref{eqn:basics2} and \eqref{eqn:basics4}; see Sec.~\ref{sec:vanishingProjectorCombinations}. 

A cyclic summation of $C^{(mn)}_{\alpha;[\beta\gamma]}$ is proportional to the real part of the cyclic summation of the torsion tensor $T^{mn}_{\alpha;\beta\gamma}$, which relates the torsion tensor to the circular shift current and offers a novel path for quantized non-linear optical responses \cite{Wojciech2024, jankowski2024non}. 

\subsubsection{Interband Wilson loop and shift vector}
We present the connection between the projector formalism and the gauge-invariant Pancharatnam-Berry phase over an interband loop, which has recently been introduced in the shift vector description of the shift current response \cite{Shi2021, Wang2022, Zhu2024}. The {\it interband Wilson loop} reads
\begin{align}
    \label{eqn:interBandWilson}
    W^{mn}_{\alpha\beta}(\bk,\bq) = \text{tr}\big[\hat e^{nm}_\beta(\bk)\,\hat e^{mn}_\alpha(\bk+\bq)\big] \, ,
\end{align}
involving the transition dipole moments in Eq.~\eqref{eqn:emna} at separated momenta. Note that the interband Wilson loop vanishes for $n=m$. It reduces to the established definition \cite{Wang2022} for non-degenerate bands; see Appendix~\ref{sec:AppendixWilsonShift}. Expanding in small $\bq$ leads to the two-state quantum geometric tensor and quantum geometric connection, 
\begin{align}
    W^{mn}_{\alpha\beta}(\bk,\bq) = -Q^{mn}_{\beta\alpha}(\bk)-\sum_\gamma q_\gamma\, C^{mn}_{\gamma;\beta\alpha}(\bk) + \cdots \, .
\end{align}
Using Eq.~\eqref{eqn:interBandWilson}, we derive the shift vector obtaining 
\begin{align}
    \label{eqn:shiftvector}
    R^{mn}_{\alpha\beta}(\bk) \equiv i \lim_{q_a\rightarrow\, 0} \partial_{q_a} \ln W_{\beta\beta}^{mn}(\bk,\bq) = i\frac{C^{mn}_{\alpha;\beta\beta}(\bk)}{Q^{mn}_{\beta\beta}(\bk)} \, ,
\end{align}
which reduces to the well-known expression \cite{Sipe2000, Shi2021, Wang2022, Ahn2022, Zhu2024} for non-degenerate bands; see Appendix~\ref{sec:AppendixWilsonShift}. In particular, we have  
\begin{align}
    &\text{Re}\, R^{mn}_{\alpha\beta} = -\frac{\text{Im}\, C^{mn}_{\alpha;\beta\beta}}{g^{mn}_{\beta\beta}} \, ,\\
    &\text{Im}\, R^{mn}_{\alpha\beta} = \frac{\text{Re}\, C^{mn}_{\alpha;\beta\beta}}{g^{mn}_{\beta\beta}} \, ,
\end{align}
which offers the interpretation of the shift vector as the metric-normalized quantum geometric connection. 

\subsubsection{Non-Abelian quantum geometric tensor} \label{sec:non-abelian}

A single $m$-degenerate state is represented by a projector $\hat P$ of rank $m$. While one can still define the quantum geometric tensor according to Eq. \eqref{eqn:Qalpha_ab}, it does not exhaust all geometric information contained in the projector. A convenient way of representing the additional information is by considering its non-Abelian version,
\begin{align}
    \mathcal{Q}^{ab}_{\alpha \beta} = \langle u_a|\big(\partial_\alpha \hat P\big)\big(\partial_\beta \hat P) | u_b \rangle \, 
\end{align}
with Bloch states $|u_a\rangle$ of the $m$-dimensional subspace. We denote this $m \times m$ matrix for fixed $\alpha, \beta$ as $\mathcal{Q}_{\alpha \beta}$ suppressing the internal indices. The Abelian quantum geometric tensor in Eq.~\eqref{eqn:Qalpha_ab} is recovered by tracing its non-Abelian version, i.e., $Q_{\alpha\beta} = \text{tr}\,\mathcal{Q}_{\alpha\beta}$. 

The antisymmetric part of $\mathcal{Q}_{\alpha\beta}$ relates to the non-Abelian Berry curvature 
\begin{align}
F_{\alpha\beta} & = i \left( \mathcal{Q}_{\alpha\beta} - \mathcal{Q}_{\beta\alpha} \right) \\ & = \partial_\alpha A_\beta - \partial_\beta A_\alpha - [A_\alpha,A_\beta] \,,
\end{align}
where $A_\alpha$ is the non-Abelian Berry connection, also known as the Wilczek-Zee connection. It is important to point out that for $m>1$, $A_{\alpha}$ cannot be recovered from $F_{\alpha\beta}$. We note that the trace of $F_{\alpha\beta}$ reduces to the Abelian Berry curvature given in Eq.~\eqref{eqn:Berrycurvature}. The symmetric part gives a matrix-valued generalization of the quantum metric,
\begin{align}
G_{\alpha\beta} = \frac{1}{2} \left( \mathcal{Q}_{\alpha\beta} + \mathcal{Q}_{\beta\alpha} \right) \, ,
\end{align}
which is a novel intrinsic geometric invariant that, among other things, leads to a series of Finsler metrics 
\begin{align}\label{eq:finsler}
g^{(n)}_{\alpha_1\beta_1\cdots\alpha_n\beta_n} = \text{tr}\big[G_{(\alpha_1\beta_1} \cdots G_{\alpha_n\beta_n)}\big] \,,
\end{align}
for $n = 1,\dots, m$ with $g^{(1)}_{\alpha \beta}$ being the quantum metric given in Eq. \eqref{eqn:quantummetric}. For $n > 1$, Eq. \eqref{eq:finsler} yields independent ways of measuring the distance between degenerate states, corresponding to Jordan's principle angles. It was shown in Ref.~\onlinecite{Avdoshkin2024} by one of us that $A_{\alpha}$ and $G_{\alpha \beta}$ give a full characterization of the intrinsic geometry of rank-$m$ projectors. Physical observables will be expressed via traces or other scaler functions of these matrix-valued objects.

\subsection{Useful projector identities}
\label{sec:UsefulProjectorIdentities}

An efficient evaluation of geometric invariants involving multiple projectors and their derivatives require a set of identities, which are implied by the four basic projector properties given in Eqs.~\eqref{eqn:basics1} to \eqref{eqn:basics4}. We focus on those identities relevant for the evaluations throughout this paper and the joint submission \cite{prl2024}. Further generalizations are straightforwardly obtained. We omit the momentum dependence throughout this section, i.e., $\hat P\equiv \hat P(\bk)$. 

\subsubsection{Trace manipulations and complex conjugation}

The cyclic property of the trace allows the cyclic permutation of the involved projectors, 
\begin{align}
    \label{eqn:cyclic}
    \text{tr}\big[\hat P_1\,\hat P_2 \,...\, \hat P_N\big] = \text{tr}\big[\hat P_2 \,...\, \hat P_N \,\hat P_1\big] \, ,
\end{align}
and further operators such as the Bloch Hamiltonian. The invariance of the trace under transposition and the hermiticity of the projectors \eqref{eqn:basics1} lead to
\begin{align}
    \label{eqn:transpose}
    \overline{\text{tr}\big[\hat P_1\,\hat P_2 \,...\, \hat P_N\big]} = \text{tr}\big[\hat P_N \,...\, \hat P_2\,\hat P_1\big] \, ,
\end{align}
where the overline denotes complex conjugation. We note that complex conjugation effectively reverses the order of the projectors under the trace. The same identity holds when projector derivatives and other Hermitian operators are involved. The two trace manipulations \eqref{eqn:cyclic} and \eqref{eqn:transpose} are essential in the evaluation of diagrammatic expansions; see, e.g., \cite{Mitscherling2018} and Sec.~\ref{sec:ShiftCurrent}. 

\subsubsection{Vanishing projector combinations}
\label{sec:vanishingProjectorCombinations}

The idempotence and orthogonality of the projectors formalized in the identity \eqref{eqn:basics2} directly implies
\begin{align}
    \label{eqn:ProjDeriv}
    \partial_\alpha \hat P = \hat P \,(\partial_\alpha \hat P) + (\partial_\alpha \hat P)\, \hat P \, .
\end{align}
Note that the projector and its derivative do not commute in general. This identity implies the vanishing of the following three combinations of projector and their derivatives,
\begin{align}
    \label{eqn:Pzero1}
    & \hat P \,(\partial_\alpha \hat P)\,\hat P = 0 \, , \\ 
    \label{eqn:Pzero2}
    & \hat P\, (\partial_\alpha \hat P) \,(\partial_\beta \hat P)\,(\partial_\gamma \hat P)\, \hat P = 0 \, , \\
    \label{eqn:Pzero3}
    & \text{tr}\big[(\partial_\alpha \hat P) \,(\partial_\beta \hat P)\,(\partial_\gamma \hat P)\big] = 0 \, .
\end{align}
Note that the last identity is only valid under the trace. The upper identities are proven by inserting \eqref{eqn:ProjDeriv} repetitively for all derivatives, which allows for a systematic derivation of further vanishing projector identities beyond the presented ones. Projector identities with higher-order derivatives can be derived using $\partial_{\alpha_1}...\,\partial_{\alpha_n} \hat P = \partial_{\alpha_n}...\,\partial_{\alpha_n} \hat P^2$. Extending the derivation to projectors of different bands, it immediately follows from \eqref{eqn:basics2} that 
\begin{align}
    \label{eqn:vanishingPmPlPn}
    \hat P_m\big(\partial_\alpha \hat P_l \big)\hat P_n = 0 
\end{align}
for $l\neq n,m$. 

\subsubsection{Derivatives of the Bloch Hamiltonian}
\label{sec:DerivativesBlochHamiltonian}

In the derivation of response functions, derivatives of the Bloch Hamiltonian are usually present. We summarize the most important identities arising from projector property \eqref{eqn:basics3}, or equivalently, 
\begin{align}
    \label{eqn:Hexpansion}
    \hat H = \sum_n \, E_n \,\hat P_n
\end{align}
where we omit the momentum dependence of the band dispersions $E_n\equiv E_n(\bk)$ and the corresponding band projectors $\hat P_n\equiv \hat P_n(\bk)$ for shorter notation. We obtain
\begin{align}
    \hat P_m\big(\partial_\alpha \hat H\big)\hat P_n = \delta_{mn}(\partial_\alpha E_m)\hat P_m\!-\epsilon_{mn}\hat P_m (\partial_\alpha \hat P_n)\hat P_n
    \label{eqn:Hdecomp}
\end{align}
with direct band gaps $\epsilon_{mn}\equiv E_m-E_n$, where we used Eq.~\eqref{eqn:vanishingPmPlPn} and $\hat P_m(\partial_\alpha \hat P_m)\hat P_n = -\hat P_m(\partial_\alpha \hat P_n)\hat P_n$. Similarly, we obtain the decomposition of the second-order Hamiltonian derivative 
\begin{align}
    \label{eqn:projIdentity3}
    \hat P_m \big(\partial_\alpha\partial_\beta \hat H\big)\hat P_n &=\,\frac{1}{2}\epsilon_{nm}\,\hat P_m\big(\partial_\alpha\partial_\beta \hat P_n\big)\hat P_n \nonumber\\&+\big(\partial_\alpha \epsilon_{nm}\big)\,\hat P_m\big(\partial_\beta \hat P_n\big)\hat P_n \nonumber  \\[2mm]
    & -E_m\,\hat P_m\big(\partial_\alpha \hat P_m\big)\big(\partial_\beta \hat P_n\big)\hat P_n \nonumber  \\[2mm]
    &-\!\!\sum_{l\neq n,m} E_l\,\hat P_m\big(\partial_\alpha \hat P_l\big)\,\hat P_l\,\big(\partial_\beta \hat P_n\big)\hat P_n \nonumber  \\ &+ (\alpha\leftrightarrow \beta) \, 
\end{align}
for $n\neq m$, where the symmetry in $\alpha$ and $\beta$ are denoted in the last line. 

\subsubsection{Two-band systems}

We have a closer look at the case of a two-band system with non-degenerate bands, which are highly constraint via the completeness relation \eqref{eqn:basics4}, i.e.,
\begin{align}
    \hat P^{}_++\hat P^{}_-=\mathds{1}_2 \, .
\end{align}
where we denote the two band projectors as $\hat P_\pm$. This relation implies $\partial^{}_\alpha \hat P^{}_+ = -\partial^{}_\alpha \hat P^{}_-$ and, thus, greatly simplifies the geometric invariants. In particular, we have
\begin{align}
    &g^{+-}_{\alpha\beta} = - g^-_{\alpha\beta} = - g^+_{\alpha\beta} = g^{-+}_{\alpha\beta} \, , \\
    &\Omega^{+-}_{\alpha\beta} = - \Omega^-_{\alpha\beta} = \Omega^+_{\alpha\beta} = -\Omega^{-+}_{\alpha\beta} \, ,
\end{align}
relating the quantum metric, Berry curvature, and their two-state variants of the two bands. We see that only a single quantum metric and Berry curvature determine all quantities. 

Similarly, we obtain the relation between the different components of the quantum geometric connections
\begin{align}
    \label{eqn:Cmn_twoband}
    C^{+-}_{\alpha;\beta\gamma}=-Q^-_{\beta;\alpha\gamma} = -\overline{Q^+_{\beta;\alpha\gamma}} = \overline{C^{-+}_{\alpha;\beta\gamma}}
\end{align}
where the overline denotes complex conjugation. These relations imply that the torsion of a two-band system vanishes, i.e.,
\begin{align}
    \label{eqn:vanishing_torsion}
    T^{+-}_{\beta;\alpha\gamma} = C^{+-}_{\gamma;\beta\alpha}-C^{+-}_{\alpha;\beta\gamma} = - Q^-_{\beta;\gamma\alpha}+Q^-_{\beta;\alpha\gamma} = 0 \, .
\end{align}
We can generalize these insights to any separation of the entire system into two subsystems, for instance, the set of occupied and unoccupied states characterized by $\hat P_\text{occ}$ and $\hat P_\text{unocc} = \mathds{1}_N-\hat P_\text{occ}$. 

Note that the relations above can also be seen as a consequence of Eqs.~\eqref{eqn:sum_Qmn} and \eqref{eqn:sum_CmnAppendix} for a two-band system. However, this does not generalize to arbitrary subsets. As we discuss in the joint submission \cite{prl2024}, the summed-over two-state quantum geometric connection $\sum_{\underset{m\in \text{unocc}}{n\in \text{occ}}} C^{mn}_{\alpha\beta} $  does not, in general, reduce to a ground state property, i.e., cannot that can be expressed using $\hat P_\text{occ}$ only.

\section{Applications of the formalism}
\label{sec:Applications}

We present two applications of the formalism described in Sec.~\ref{sec:projectors}, which both provide further physical insights into the quantum geometric invariants and illustrate how derivations of observables and their relation to geometric invariants are performed within the projector calculus. We start with the quantum state geometry of polarization in Sec.~\ref{sec:Polarization}, which focuses on the single-state quantities, and close by the derivation of the injection and shift current in Sec.~\ref{sec:ShiftCurrent}, which elaborates on the two-state quantities. 

\subsection{The quantum state geometry of polarization}
\label{sec:Polarization}

In the first example, we focus on the polarization of extended periodic systems. We start by introducing the necessary concepts and formalism. The approach is known as the modern theory of polarization in the literature and was motivated by the need to define spontaneous polarization in ferroelectrics rigorously \cite{Kingsmith1993, Resta2007}. The main challenge is that the naive definition of the polarization vector $\bP$ via 
\begin{align}
    \bP = \int_{\text{sample}} \hspace{-8mm}d^d\bx \,\,\,~ n(\bx) ~ \bx
\end{align}
does not yield a well-defined bulk property due to corrections of the order of the sample volume $\mathcal{O}(V_\text{sample})$ coming from adding charges at the boundary. One way to properly define bulk polarization is via a sum over Wannier functions $|W_n\rangle$ constructed from the occupied bands
\begin{align}
    \bP = \sum_{n\in \text{occ}} \langle W_n|\,\hat \bx\,|W_n\rangle \, ,
\end{align}
with summation over the Wannier functions spanning all occupied bands. In the following, we take an alternative approach to deriving the average polarization and higher-order moments of the polarization distribution. For this, let us define the total (many-body) position operator on a lattice as $\hat{\bX} = \sum_{i\in\text{sites}} \hat{\bx}_i$, which is related to the polarization operator by $\hat \bP = -e \,\hat \bX$. The generating function for its moments reads
\begin{align}
    C(\bq) &= \langle \Psi | e^{i \bq\cdot\hat{\bX}} | \Psi \rangle \nonumber\\&= 1 + i \sum_\alpha q_{\alpha} \langle \Psi | \hat{X}_{\alpha} | \Psi \rangle \nonumber\\&- \frac{1}{2}\sum_{\alpha,\beta} q_{\alpha} q_{\beta} \langle \Psi | \hat{X}_{\alpha} \hat{X}_{\beta} | \Psi \rangle_c\\
    &+ \frac{1}{6}\sum_{\alpha,\beta,\gamma} q_{\alpha} q_{\beta} q_{\gamma} \langle \Psi | \hat{X}_{\alpha} \hat{X}_{\beta} \hat{X}_{\gamma} | \Psi \rangle_c + \dots\,,
\end{align}
where $|\Psi\rangle$ is the wave function of the electronic state and the subscript $c$ stands for the connected correlator, which reads for the second and third cumulant, 
\begin{align}
    &\langle \hat{X}_{\alpha} \hat{X}_{\beta} \rangle_c = \langle \hat{X}_{\alpha} \hat{X}_{\beta} \rangle - \langle \hat{X}_{\alpha} \rangle \langle \hat{X}_{\beta} \rangle \,,\\
    &\langle \hat{X}_{\alpha} \hat{X}_{\beta} \hat{X}_{\gamma} \rangle_c = \langle \hat{X}_{\alpha} \hat{X}_{\beta} \hat{X}_{\gamma} \rangle \nonumber \\ &\hspace{10mm}- \langle \hat{X}_{\alpha} \hat{X}_{\beta} \rangle \langle \hat{X}_{\gamma} \rangle- \langle \hat{X}_{\beta} \hat{X}_{\gamma} \rangle \langle \hat{X}_{\alpha} \rangle - \langle \hat{X}_{\gamma} \hat{X}_{\alpha} \rangle \langle \hat{X}_{\beta}\rangle \nonumber\\
    &\hspace{10mm}+ 2 \langle \hat{X}_{\alpha} \rangle  \langle \hat{X}_{\beta} \rangle  \langle \hat{X}_{\gamma} \rangle\,.
\end{align}
In the non-interaction case, $|\Psi\rangle  = \prod_{n,\bk} u_n(\bk) \,\hat c^{\dagger}_{n\bk}|0 \rangle$ is given by the Slater determinant constructed from the occupied states. As discussed in the following subsection, Slater determinants have the remarkable property of only depending on the subspace spanned by the participating states. We will develop an approach to describing the center-of-mass information in the Slater determinant in terms of the projector on all occupied bands, $\hat P_\text{occ}(\bk) = \sum_{n \in \text{occ}} \hat P_n(\bk)$. 

\subsubsection{Slater determinants and the Pl\"ucker map} \label{sec:pluecker}

The Slater determinant is a natural way of constructing a many-body fermionic state out of a collection of $m$ single-body states $|\psi_i \rangle$ via
\begin{align}
    &\Psi(x_1, x_2, x_3, \cdots) \nonumber\\&= \frac{1}{m!}\,\det\!\begin{pmatrix} \psi_1(x_1) & \psi_2(x_1) & \psi_3(x_1)  & \cdots \\
    \psi_1(x_2) & \psi_2(x_2) & \psi_3(x_2)  & \cdots \\
    \psi_1(x_3) & \psi_2(x_3) & \psi_3(x_3) & \cdots \\
    \vdots & \vdots & \vdots & \ddots 
\end{pmatrix}\,.
\end{align}
Due to the properties of the determinant, the set of states $U |\psi_i \rangle$, where $U$ is a unitary, gives the same state modulo a phase. Thus, the resulting state is a function of $\text{span}\big[| \psi_i\rangle\big]$ only. This construction is known as the Pl\"ucker map in mathematics. Formally, it is a map from the Grassmanian $Gr(m, V)$, the space of $m$-planes in $V$, to the projectivization of the exterior power of the original vector space $P(\Lambda^m V)$. In physical terms, the exterior power $\Lambda^m V$ is the many-body fermionic Hilbert space. Explicitly, the Pl\"ucker map is given by
\begin{align}
    \text{span}(w_1, \dots, w_m) \to \left[ w_1 \wedge \dots \wedge w_m\right]\,,
\end{align}
where $\wedge$ stands for the exterior product and $[...]$ stands for projectivization. We use this map to define the geometric objects associated with a single state of an entire subspace. The starting point is the inner product that is induced on the exterior power
\begin{align}
    \langle \Lambda v, \Lambda w \rangle = \det \big( \langle v_i| w_j\rangle \big).
\end{align}
with the matrix $\big( \langle v_i| w_j\rangle \big)$ involving all combinations of the single-particle states within the many-body state $\Lambda v$ and $\Lambda w$. With that, given three $m$-dimensional subspaces spanned by $|v_i\rangle$, $|w_i\rangle$ and $|u_i\rangle$ we define the three-point function \cite{Avdoshkin2022} via
\begin{align} \label{eq:3pt_function}
    &\langle \Lambda v, \Lambda w \rangle 
    \langle \Lambda w, \Lambda u \rangle 
    \langle \Lambda u, \Lambda v \rangle \nonumber\\&= \det \big(\langle v_i | w_j \rangle\big) \,\det \big(\langle w_i | u_j \rangle\big) \, \det \big(\langle u_i | v_j \rangle\big) \nonumber\\&= \sum_{s,s'}\det \big(\langle v_i | w_s \rangle\langle w_s | u_{s'} \rangle\langle u_{s'} | v_j \rangle\big) \, ,
\end{align}
where we used $\det(AB) = \det A \det B$. We can alternatively write it as
\begin{align}
    \text{det}_{v} \big[\hat P_v \, \hat P_w \, \hat P_u \, \hat P_v \big]\, ,
\end{align}
where $\hat P_v = \sum_s |v_s\rangle \langle v_s|$ (and analogously for $w$ and $u$) and $\text{det}_v$ is taken over the $\hat P_v$ subspace. Considering that $\{v_s\}$ forms a basis of the image $\text{Im} \,\hat P_v$, we need to compute $\langle v_s| \hat P(w) \hat P(u)| v_{s'} \rangle$. We keep the last projector to make the projection onto the subspace explicit. The described procedure allows us to generalize the geometric characterization of the $\mathbb{C}P^n$ theory as presented in Ref.~\onlinecite{Avdoshkin2022} by one of us to invariants that can be written as
\begin{align}
    \text{det}_{v_1} \big[\hat P_{v_1} \,\hat P_{v_2} \,\cdots \, \hat P_{v_m} \, \hat P_{v_1} \big] \, .
\end{align}
In particular, all invariants of that form can be reduced to 3-point functions such as given in Eq.~\eqref{eq:3pt_function}. 

\subsubsection{Polarization}

Consider an insulator with $N_\text{occ}$ filled bands. The corresponding ground state wavefunction is $|\Psi\rangle = \prod_{n,\bk} u_n(\bk) \,\hat c^{\dagger}_{n\bk}|0 \rangle$ with vacuum $|0\rangle$ and fermionic creation operators $\hat c^\dagger_{n\bk}$. We notice that $e^{i \bq \cdot \hat \bX} | \Psi \rangle = \prod_{n,\bk} u_n(\bk) \,\hat c^{\dagger}_{n,\bk-\textbf{q}}|0 \rangle$. Thus, the generating function reads \cite{Patankar2018}
\begin{align} 
    \label{eq:C_of_q}
    C(\bq) &\equiv 
    \int_\text{BZ} \log \det \Big( \langle u_i(\bk)|u_j(\bk+\bq)\rangle  \Big)_{ij} \,,
\end{align}
where the determinant is taken over the matrix with entries $\langle u_i(\bk)|u_j(\bk+\bq)\rangle$. In Appendix~\ref{sec:PolarizationAppendix}, we redo the derivation from Ref.~\onlinecite{Avdoshkin2022} for multiple filled bands. The resulting  expression is
\begin{align} \label{eq:C_q_expr}
    \frac{\log C(\bq)}{V} = \!\sum_\alpha q^\alpha\mathcal{A}_\alpha\!+\!\sum_\alpha q^\alpha \!\!\int_\text{BZ} \int_0^1\!\!\!\!dt\,\, \mathcal{A}_\alpha^{\bk}(\bk+\bq t) 
\end{align}
with the vector-valued 2-point function
\begin{align}
    \label{eqn:Aabkmaintext}
    \mathcal{A}_{\alpha}^{\bk}(\bk') = \text{tr}\Big[\hat P_{\bk}\,\big(\hat P_{\bk} \, \hat P_{\bk'}\,\hat P_{\bk}\big)^{-1} \,\hat P_{\bk}\,\big(\partial_{\alpha}\,\hat P_{\bk'}\big) \,\hat P_{\bk'} \Big] \, .
\end{align}
This result generalizes Eq.~(17) in Ref.~\onlinecite{Avdoshkin2022} to higher-rank projectors and was reported in the companion article \cite{prl2024}. The inversion operation $(\cdot)^{-1}$ is performed over the $\hat P_{\bk}$ subspace. In the non-degenerate case, the product $\hat P_\bk \hat P_{\bk'} \hat P_\bk$ is a rank-1 object so that the inverse is taken over a number. This number can be pulled out of the trace. Combining the rest in the trace and rewriting the number as a trace over the full space, we recover the rank-1 formula introduced in Ref.~\onlinecite{Avdoshkin2022} up to complex conjugation, which we redefined for the convenience of a simpler notation. In the degenerate case, we can evaluate Eq.~\eqref{eqn:Aabkmaintext} by interpreting the objects as matrices within a particular basis. In particular, all operations are taken over the degenerate subspace only, or the inverse can be expanded to the size of the original case after inversion restricted to the subspace by adding zero elements in the matrices.

The generating function in Eq. \eqref{eq:C_q_expr} allows us to derive the cumulants of polarization through the expansion of $\mathcal{A}_{\alpha}^{\bk}(\bk+\bq t)$, that is,
\begin{align}
    \mathcal{A}_\alpha^{\bk}(\bk+\bq) &= \sum_\beta Q_{\alpha\beta}(\bk) \,q^\beta \nonumber \\ &+ \sum_{\beta,\gamma} Q_{\alpha;\beta\gamma}(\bk)\, q^\beta q^\gamma + \cdots \, ,
\end{align}
where we identify the quantum geometric tensor and quantum geometric connection given in Eqs.~\eqref{eqn:Qalpha_ab} and \eqref{eqn:Qalphabetagamma} with respect to the projector onto the $N_\text{occ}$ filled bands for the linear and quadratic term of the expansion, respectively. This leads to the explicit expressions for the second and third cumulants of the polarization distribution,
\begin{align}\label{eq:q2}
    \langle X_{\alpha} X_{\beta} \rangle_c &= V\!\!\! \int_\text{BZ}\!\!\!\text{Re}\, Q_{\alpha\beta}(\bk) \, ,\\
    \langle X_{\alpha} X_{\beta} X_{\gamma} \rangle_c &= V\!\!\! \int_\text{BZ}\!\!\!\text{Im}\, Q_{\alpha;\beta\gamma}(\bk)\label{eq:q3} \, ,
\end{align}
as reported in the companion article \cite{prl2024}. We note that higher cumulants are, in general, not simply related to the single-state (non-degenerate) geometric invariants given in Eq.~\eqref{eqn:localGeoInv} by replacing the non-degenerate with the degenerate projectors. Instead, we expect deviations in the expansion of $\mathcal{A}_\alpha^{\bk}(\bk+\bq)$ between the non-degenerate and degenerate cases starting in the third order.


\subsection{The injection and shift currents}
\label{sec:ShiftCurrent}

We demonstrate the projector calculus for injection and shift currents, where the relevance of higher-order multi-state quantum geometry, that is, the two-state quantum geometric tensor and the two-state quantum geometric connection, is known \cite{Holder2020, Ahn2022, Kaplan2023}. We explicitly show that the gauge-invariance of the projector calculus allows for a straightforward and transparent derivation despite the need to include various contributions that arise within the standard diagrammatic expansion. Within the projector formalism, unnecessary separations into gauge-dependent terms are avoided, and the expressions are provided in a unified language that allows for simpler recombination of contributions from different diagrams.

We consider the second-order DC optical response 
\begin{align}
    \label{eqn:definitionSigma}
    j^a(0;\omega,-\omega) = \sum_{b,c}\sigma^{a;bc}(0;\omega,-\omega)\,\cE^b(\omega)\,\cE^c(-\omega),
\end{align}
where $E^a(\omega)$ is the Fourier component of the external electric fields and $j^a$ is the electric current. We will use the results of Ref.~\onlinecite{Holder2020} which expressed $\sigma^{a;bc}$ in terms of the Bloch Hamiltonian $\hat{H}$. For a structured calculation, we split the conductivity into three parts distinct by the maximal order of the vertices from three to one, defining
\begin{align}
    \label{eqn:sigmaTensor}
    \sigma^{\aCord;\bCord\cCord}(\omega) &\equiv \sigma^{\aCord;\bCord\cCord}(0,\omega,-\omega)\nonumber\\&= \frac{e^3}{\hbar^2 \omega^2}  \!\!\int_\text{BZ}\!\Big(\sigma^{\aCord;\bCord\cCord}_{(0)} + \sigma^{\aCord;\bCord\cCord
    }_{(1)}(\omega) + \sigma^{\aCord;\bCord\cCord}_{(2)}(\omega) \Big) \,.
\end{align}
For shorter notation, we do not explicitly write the momentum dependence within the integrand of the integral over the Brillouin zone (BZ). We focus on the contribution
\begin{align}
    \label{eqn:sigma2}
    \sigma_{(2)}^{\aCord;\bCord\cCord}(\omega) &= \text{tr}\Bigg[ \hat P_\text{occ}\Bigg(
    \bigg[\bigg[\partial_\cCord \hat H\,,\, \frac{\partial_\aCord \hat H}{ \epsilon - i \gamma}\bigg]\,,\, \frac{\partial_\bCord \hat H}{\omega + \epsilon + i \Gamma}\bigg] \nonumber \\&\hspace{4mm}+ \bigg[\bigg[\partial_\bCord \hat H\,,\, \frac{\partial_\aCord \hat H}{ \epsilon - i \gamma}\bigg]\,,\, \frac{\partial_\cCord \hat H}{-\omega + \epsilon + i \Gamma}\bigg] \Bigg)\Bigg] \, ,  
\end{align}
involving only first-order vertices $\partial_a\hat H$ and present more details for the other two contributions in Appendix~\ref{sec:shiftcurrentAppendix}. The individual terms are defined within the band basis, for instance, $\langle u_n|\partial_\aCord \hat H/(-\epsilon+i\gamma)|u_m\rangle=\langle u_n|\partial_\aCord \hat H|u_m\rangle/(-\epsilon_{nm}+i\gamma)$. We refer to Refs.~\cite{Holder2020, Jiang2025} for a detailed physical motivation of the intraband relaxation rate $\gamma$ and interband relaxation rate $\Gamma$. The numbers $\omega$, $\gamma$, and $\Gamma$ are understood to be multiplied with the identity matrix. $\epsilon_{nm}=E_n-E_m$ is the direct band gap between bands $n$ and $m$. The $[\cdot,\cdot]$ is the commutator of the involved matrices. We simplified the notation given in Ref.~\onlinecite{Holder2020} by introducing 
\begin{align}
    \label{eqn:PoccFermi}
    \hat P_\text{occ} \equiv \sum_n f_n \,\hat P_n
\end{align}
with Fermi function $f_n$ and orthogonal band projector $\hat P_n$ with corresponding (non-)degenerate band energy $E_n$. Note that $\hat P_\text{occ}$ is {\it not} a projector expect for a band insulator at zero temperature but this close connection motivates us to introduce the notation. We do not need to assume that $\hat P_\text{occ}$ is a projector throughout the following derivation. 

\subsubsection{General strategy}

Before performing the derivation in detail, we sketch the main steps, which we anticipate to apply to other observables as well, as long as the starting point is similar. We provided a detailed discussion on anticipated further applications of the formalism within the companion article \cite{prl2024}. Within the following derivation, we aim to separate quantum geometric invariants that only involve projectors as introduced in Sec.~\ref{sec:projectors} from a remaining part. This remaining spectral part usually captures the information about the band energies, temperature, or other external parameters. We perform this separation in an iterative process involving three main steps, which we illustrate for the injection and shift current in the following.

{\bf (Step 1)} In order to evaluate the expression in Eq.~\eqref{eqn:sigma2}, we first aim to separate the contributions involving spectral information such as the band energies, referred to as {\it spectral contribution}, from (preliminary) geometric contributions that might still contain spectral information. For this, we introduce projector expressions via Eqs.~\eqref{eqn:basics4} and \eqref{eqn:Hexpansion}. It might be convenient to keep derivatives of the Hamiltonian explicit in order to avoid, at that point of the derivation, lengthy expressions arising from identities such as those given in Eqs.~\eqref{eqn:Hdecomp} and \eqref{eqn:projIdentity3}.

{\bf (Step 2)} We use projector identities such as those presented in Sec.~\ref{sec:UsefulProjectorIdentities} to simplify the individual (preliminary) geometric contributions, for instance, when performing the commutators. This generically leads to constraints for the band summations, further simplifying the spectral contribution. 

{\bf (Step 3) }We conclude by regrouping contributions with the same spectral contributions after the simplifications of the second step. Usually, further projector identities, as presented in Sec.~\ref{sec:UsefulProjectorIdentities}, can be applied to obtain the final form, where we finally identify the relevant geometric invariant that only contains band projectors. In the following application to the injection and shift current, we know that the final result will involve only the quantities presented in Sec.~\ref{sec:ImportantGeometricInvariants}. In general, novel geometric invariants may be necessary to consider.

\subsubsection{Step-by-step evaluation of Eq.~\eqref{eqn:sigma2}}

We notice that it is sufficient to evaluate 
\begin{align}
    \label{eqn:Bexpression}
    &B^{\aCord\bCord\cCord}(\omega) \equiv \text{tr}\Bigg[\hat P_\text{occ}\bigg[\bigg[\partial_\aCord\hat H,\frac{\partial_\bCord \hat H}{\epsilon-i\gamma}\bigg],\frac{\partial_\cCord\hat H}{\omega+\epsilon+i\Gamma}\bigg]\Bigg] \, ,
\end{align}
since $\sigma^{\aCord;\bCord\cCord}_{(2)}(\omega) = B^{\cCord\aCord\bCord}(\omega)+B^{\bCord\aCord\cCord}(-\omega)$. Following the first step of the general strategy, we focus on the two fractions involving the band energies. Using Eq.~\eqref{eqn:Hexpansion}, we see that the contribution involving the interband relaxation rate $\Gamma$ decomposes into
\begin{align}
    \label{eqn:Gamma_expansionMain}
    \frac{\partial_\aCord\hat H}{\omega+\epsilon+i\Gamma} 
    &= \sum_{n}\frac{\partial_\aCord E_n }{\omega+i\Gamma} \hat P_n \nonumber \\ &- \sum_{\underset{n\neq m}{n,m}}\frac{\epsilon_{nm}}{\omega+\epsilon_{nm}+i\Gamma} \, \hat P_n(\partial_\aCord \hat P_m)\hat P_m \, ,
\end{align}
where we used Eq.~\eqref{eqn:Hdecomp} to evaluate the derivative of the Hamiltonian. Equivalently, we apply the same decomposition for the contribution involving the intraband relaxation rate $\gamma$ and obtain
\begin{align}
    \label{eqn:gamma_expansionMain}
    &\frac{\partial_\aCord \hat H}{\epsilon-i\gamma} 
    = \frac{i}{\gamma}\sum_{n}(\partial_\aCord E_n)\hat P_n - \sum_{n}\hat Q^{(\gamma)}_n(\partial_\aCord \hat P_n)\hat P_n \, , 
\end{align}
where we introduce
\begin{align}
    \label{eqn:Qg_expansion}
    \hat Q^{(\gamma)}_n \equiv \sum_{m\neq n} \frac{\epsilon_{mn}}{\epsilon_{mn}-i\gamma}\hat P_m \approx \mathds{1}_N - \hat P_n
    \, ,
\end{align}
which projects onto the complement of band $n$ in our limit of interest $\gamma\ll |\epsilon_{mn}|$. We see that we obtain four contributions when inserting Eq.~\eqref{eqn:Gamma_expansionMain} and \eqref{eqn:gamma_expansionMain} into Eq.~\eqref{eqn:Bexpression}. We analyze the four contributions in the following. We notice that the four contributions involve two structures that simplify to 
\begin{align}
    \label{eqn:Pocc_identity1Main}
    & \text{tr}\Big[\hat P_\text{occ}\big[\hat O, \hat P_n\big]\Big] 
    = 0 \, , \\
    \label{eqn:Pocc_identity2Main}
    &\text{tr}\Big[\hat P_\text{occ}\big[\hat O_1\,,\hat P_n \,\hat O_2 \hat P_m\big]\Big] 
    = -f_{nm}\,\text{tr}\Big[\hat O_1 \hat P_n \,\hat O_2 \hat P_m\Big] \, ,
\end{align}
using Eq.~\eqref{eqn:PoccFermi}, which hold for arbitrary operators $\hat O$, $\hat O_1$, and $\hat O_2$. 

Following the second step of our general strategy, we start with the term involving the first terms of Eq.~\eqref{eqn:Gamma_expansionMain} in combination with the first and second term of Eq.~\eqref{eqn:gamma_expansionMain}. Using the identity in Eq.~\eqref{eqn:Pocc_identity1Main}, we see that both contributions vanish, i.e.,
\begin{align}
     &\text{tr}\Big[\hat P_\text{occ}\Big[\!\big[\partial_\aCord\hat H,\hat P_n\big], \hat P_{n'}  \big]\!\Big] = 0 \, ,\\
     &\text{tr}\Big[\hat P_\text{occ}\big[ \big[\partial_\aCord\hat H,\hat Q^{(\gamma)}_n(\partial_\bCord \hat P_n)\hat P_n\big], \hat P_{n'} \big]\Big] = 0 \, .
\end{align}
Combining the second term of Eq.~\eqref{eqn:Gamma_expansionMain} with the first term of Eq.~\eqref{eqn:gamma_expansionMain} involves 
\begin{align}
     &-\text{tr}\Big[\hat P_\text{occ}\big[\big[\partial_\aCord\hat H,\hat P_n\big],  \, \hat P_{n'}(\partial_\cCord \hat P_{m'})\hat P_{m'} \big]\Big] \\
     &=
     f_{n'm'}\text{tr}\Big[\big[\partial_\aCord\hat H,\hat P_n\big] \, \hat P_{n'}(\partial_\cCord \hat P_{m'})\hat P_{m'} \big]\Big]\\
     &=
     \big(\delta_{nn'}-\delta_{nm'}\big)\,f_{n'm'}\text{tr}\Big[(\partial_\aCord\hat H) \, \hat P_{n'}\,(\partial_\cCord \hat P_{m'})\hat P_{m'} \big]\Big] \, ,
\end{align}
where we have used the basic projector property \eqref{eqn:basics2} to derive the constraint on the band summations. Collecting the spectral contributions and performing the band summations leads to 
\begin{align}
     \frac{i}{\gamma}\sum_{\underset{n\neq m}{n,m}}\frac{(\partial_\bCord \epsilon_{nm})\,\epsilon_{nm}\,f_{nm}}{\omega+\epsilon_{nm}+i\Gamma}\,\text{tr}\Big[(\partial_\aCord\hat H) \, \hat P_{n}\,(\partial_\cCord \hat P_{m})\hat P_{m} \big]\Big] \, .
\end{align}
The last term arising from Eq.~\eqref{eqn:Bexpression} involves the second term of Eq.~\eqref{eqn:Gamma_expansionMain} in combination with the second term of Eq.~\eqref{eqn:gamma_expansionMain}. Using the approximation $\gamma\ll |\epsilon_{nm}|$ in Eq.~\eqref{eqn:Qg_expansion}, we obtain after similar steps
\begin{align}
    &\sum_{\underset{n\neq m}{n,m}}\frac{\epsilon_{nm}\,f_{nm}}{\omega+\epsilon_{nm}+i\Gamma}\,\text{tr}\Big[\big[(\partial_\aCord\hat H)(\partial_\bCord \hat P_n)\nonumber \\[-3mm] &\hspace{25mm}+(\partial_\bCord \hat P_m)(\partial_\aCord\hat H)\big]\hat P_n\,(\partial_\cCord \hat P_{m})\hat P_{m}\Big] \, .
\end{align}
Combining these results, we find that the contributions to the injection and shift current arising from the contributions involving three single-order vertices read 
\begin{align}
    \label{eqn:sigma2_final_main}
    &\sigma^{\aCord;\bCord\cCord}_{(2)}(\omega)     
    =\sum_{\underset{n\neq m}{n,m}} \mathfrak{\tilde s}^{(1)}_{nm}(\omega)\,\text{tr}\Big[(\partial_\cCord\hat H)\hat P_{n}(\partial_\bCord \hat P_{m})\hat P_{m}\big]\!\Big] \nonumber \\
    &\hspace{4mm}+\sum_{\underset{n\neq m}{n,m}}\mathfrak{\tilde s}^{(1)}_{nm}(-\omega)\,\text{tr}\Big[(\partial_\bCord\hat H) \hat P_{n}(\partial_\cCord \hat P_{m})\hat P_{m} \big]\!\Big] \nonumber \\
    &\hspace{4mm}-\sum_{\underset{n\neq m}{n,m}} \mathfrak{\tilde s}^{(2)}_{nm}(\omega)\,\text{tr}\Big[\big[(\partial_\cCord\hat H)(\partial_\aCord \hat P_n)+(\partial_\aCord \hat P_m)(\partial_\cCord\hat H)\big]\nonumber \\[-6mm] &\hspace{34mm}\times\hat P_n(\partial_\bCord \hat P_{m})\hat P_{m}\Big] \nonumber \\
    &\hspace{4mm}-\sum_{\underset{n\neq m}{n,m}} \mathfrak{\tilde s}^{(2)}_{nm}(-\omega)\,\text{tr}\Big[\big[(\partial_\bCord\hat H)(\partial_\aCord \hat P_n)+(\partial_\aCord \hat P_m)(\partial_\bCord\hat H)\big]\nonumber \\[-6mm] &\hspace{34mm}\times \hat P_n(\partial_\cCord \hat P_{m})\hat P_{m}\Big] \, .
\end{align}
We have introduced the short notation for the part of the spectral contributions that we have identified so far, 
\begin{align}
    &\mathfrak{\tilde s}^{(1)}_{nm}(\omega) \equiv \frac{i}{\gamma}\frac{(\partial_a \epsilon_{nm})\epsilon_{nm}f_{nm}}{\omega+\epsilon_{nm}+i\Gamma} \, , \\
    &\mathfrak{\tilde s}^{(2)}_{nm}(\omega) \equiv \frac{\epsilon_{nm}f_{nm}}{\omega+\epsilon_{nm}+i\Gamma} \, .
\end{align}
As shown in Appendix~\ref{sec:PolarizationAppendix}, the first contribution in Eq.~\eqref{eqn:sigmaTensor} vanishes, $\sigma^{\aCord;\bCord\cCord}_{(0)}=0$, to leading order in $\gamma\ll |\epsilon_{nm}|$. The contribution to the injection and shift current involving both first- and second-order vertices read
\begin{align}
    \label{eqn:sigma1_final_main}
    \sigma^{\aCord;\bCord\cCord}_{(1)}(\omega) = &-\!\sum_{\underset{n\neq m}{n,m}}\!
    \mathfrak{\tilde s}^{(2)}_{nm}(\omega) \,\text{tr}\big[(\partial_\aCord\partial_\cCord \hat H)\hat P_n(\partial_\bCord\hat P_m)\hat P_m\big] \nonumber\\&\hspace{-2mm}+\!\!\sum_{\underset{n\neq m}{n,m}}\!
     \mathfrak{\tilde s}^{(2)}_{nm}(-\omega) \,\text{tr}\big[(\partial_\aCord\partial_\bCord \hat H)\hat P_n(\partial_\cCord\hat P_m)\hat P_m\big] .
\end{align}
In light of the third step in our general strategy, we note that the contributions that are proportional to $\mathfrak{\tilde s}^{(1)}_{nm}$ cannot be further combined with other contributions. In contrast, several contributions involve $\mathfrak{\tilde s}^{(2)}_{nm}$ that we combine in the following.

\subsubsection{Combining all contributions to identify the relevant geometric invariants}

We start with the term that cannot be further reduced. Using the identity \eqref{eqn:Hdecomp} to evaluate the derivative of the Bloch Hamiltonian and further simplify the resulting combinations of projectors via our results in Sec.~\ref{sec:vanishingProjectorCombinations}, we identity the two-state quantum geometric tensor, i.e.,
\begin{align}
    &\text{tr}\Big[\big(\partial_\cCord \hat H\big)\hat P_n\big(\partial_\bCord \hat P_m\big)\hat P_m\Big] 
    = \epsilon_{nm}\,Q^{nm}_{\cCord\bCord} \, . 
\end{align}
Note that the band diagonal contribution vanishes due to $\hat P_n(\partial_b \hat P_n)\hat P_n =0$. 

The contributions in Eq.~\eqref{eqn:sigma1_final_main} can be combined with those in Eq.~\eqref{eqn:sigma2_final_main} proportional to $\mathfrak{\tilde s}^{(2)}_{nm}$. Since the second derivative of the Bloch Hamiltonian is already given in Eq.~\eqref{eqn:projIdentity3}, we focus only on the novel projector contributions in Eq.~\eqref{eqn:sigma2_final_main}. Expanding the Bloch Hamiltonian in its projector representation leads to
\begin{align}
    \label{eqn:projIdentity1}
    &\hat P_m \big(\partial_\cCord \hat H\big)\big(\partial_\aCord \hat P_n\big)\hat P_n \nonumber \\ =& \,\,\,\big(\partial_\cCord E_m\big)\hat P_m \big(\partial_\aCord \hat P_n\big)\hat P_n + E_m \,\hat P_m\big(\partial_\cCord \hat P_m\big)\big(\partial_\aCord \hat P_n\big)\hat P_n \nonumber\\&+ \sum_{l\neq n,m} E_l\,\hat P_m\big(\partial_\cCord \hat P_l\big)\,\hat P_l\,\big(\partial_\aCord \hat P_n\big)\hat P_n  \, ,
\end{align}
and a similar expression for $\hat P_m (\partial_\aCord \hat P_m)(\partial_\cCord \hat H)\hat P_n$. We note that the summations over other bands than $n$ and $m$ cancel when combining all three contributions. We find
\begin{align}
    &\text{tr}\bigg[\Big[\big(\partial_\aCord\partial_\cCord \hat H\big)+ \big(\partial_\cCord \hat H\big)\big(\partial_\aCord \hat P_n\big) + \big(\partial_\aCord \hat P_m\big)\big(\partial_\cCord \hat H\big)\Big]\nonumber \\ &\hspace{10mm}\times \hat P_n \big(\partial_\bCord \hat P_m\big)\hat P_m\bigg] 
    \nonumber \\[2mm] 
    &= \big(\partial_\aCord \epsilon_{nm}\big)\,Q^{nm}_{\cCord\bCord} + \epsilon_{nm}\,C^{mn}_{\aCord;\bCord\cCord} \, ,
\end{align}
where we identified the two-state quantum geometric tensor $Q^{nm}_{bc}$ and the two-state quantum geometric connection $C^{mn}_{a;bc}$ in combination with further spectral contributions. We recombine these spectral contributions with those obtained previously defining 
\begin{align}
    \label{eqn:s1nm}
    &\mathfrak{s}^{(1)}_{nm}(\omega) \equiv \frac{i}{\gamma}\frac{(\partial_a \epsilon_{nm})(\epsilon_{nm})^2f_{nm}}{\omega+\epsilon_{nm}+i\Gamma} \, , \\
    &\mathfrak{s}^{(2a)}_{nm}(\omega) \equiv \frac{(\partial_a \epsilon_{nm})\epsilon_{nm}f_{nm}}{\omega+\epsilon_{nm}+i\Gamma} \, , \\
    &\mathfrak{s}^{(2b)}_{nm}(\omega) \equiv \frac{(\epsilon_{nm})^2f_{nm}}{\omega+\epsilon_{nm}+i\Gamma} \, , 
    \label{eqn:s2bnm}
\end{align}
which contains all dependencies on the band dispersions, the scattering rates, and driving frequency. We conclude by combining all previous results and obtain 
\begin{align}
    \label{eqn:finalSigma}
    \sigma^{\aCord;\bCord\cCord}(\omega)    
    &=\frac{e^3}{\hbar^2\omega^2}\sum_{\underset{n\neq m}{n,m}}\int_\text{BZ}\!\!\Big(\mathfrak{s}^{(1)}_{nm}(\omega)Q^{nm}_{\cCord\bCord}+\mathfrak{s}^{(1)}_{nm}(-\omega)Q^{nm}_{\bCord\cCord}\Big)  \nonumber \\
    &-\frac{e^3}{\hbar^2\omega^2}\sum_{\underset{n\neq m}{n,m}}\int_\text{BZ}\!\!\Big(\mathfrak{s}^{(2a)}_{nm}(\omega)Q^{nm}_{\cCord\bCord} +\mathfrak{s}^{(2a)}_{nm}(-\omega)Q^{nm}_{\bCord\cCord}\Big)\nonumber \\
    &-\frac{e^3}{\hbar^2\omega^2}\sum_{\underset{n\neq m}{n,m}}\int_\text{BZ}\!\!\Big(\mathfrak{s}^{(2b)}_{nm}(\omega)C^{mn}_{\aCord;\bCord\cCord}+ \mathfrak{s}^{(2b)}_{nm}(-\omega)C^{mn}_{\aCord;\cCord\bCord} \Big)
\end{align}
in leading order in $\gamma/|\epsilon_{nm}|\ll 1$. We see the symmetry in $(\cCord,\omega)\leftrightarrow (\bCord,-\omega)$ as required by the definition given in Eq.~\eqref{eqn:definitionSigma}. Focusing on the resonant parts via the replacement $\big[x+i\Gamma\big]^{-1} \rightarrow -i\pi\,\delta(x)$, we find that the second term in Eq.~\eqref{eqn:finalSigma} vanishes. The first and third terms are identified as injection and shift current $\sigma^{\aCord;\bCord\cCord}(\omega) = \sigma^{\aCord;\bCord\cCord}_\text{inj}(\omega) + \sigma^{\aCord;\bCord\cCord}_\text{shift}(\omega)$ given as
\begin{align}
    &\sigma^{\aCord;\bCord\cCord}_\text{inj}(\omega)\!\equiv \!\frac{2\pi e^3}{\gamma\hbar^2}\!\!\sum_{\underset{n\neq m}{n,m}}\!\int_\text{BZ} \!\!\!\!\!\!\delta(\omega-\epsilon_{nm})f_{nm}(\partial_\aCord \epsilon_{nm})\,Q^{nm}_{\bCord\cCord} \!,\\
    &\sigma^{\aCord;\bCord\cCord}_\text{shift}(\omega) \!\equiv\! \frac{i \pi e^3}{\hbar^2}\!\!\sum_{\underset{n\neq m}{n,m}}\!\int_\text{BZ}\!\!\!\!\!\!\delta(\omega\!-\!\epsilon_{nm})f_{nm}\!\big(C^{mn}_{\aCord;\cCord\bCord}\!-\!C^{nm}_{\aCord;\bCord\cCord} \big).
\end{align}
These formulas agree with those given in Ref.~\onlinecite{Ahn2020}, which we generalized to degenerate bands. The intraband relaxation rate $\gamma$ within the two-lifetime model \cite{Holder2020, Jiang2025} corresponds to the relaxation time, at which the injection current eventually saturates \cite{Sipe2000, Ahn2020}. For finite interband relaxation rate $\Gamma$, a third contribution is found in Eq.~\eqref{eqn:finalSigma} as previously reported in Ref.~\onlinecite{Kaplan2023}. Setting $\Gamma=0$ in Eqs.~\eqref{eqn:s1nm} to \eqref{eqn:s2bnm} allows to determine off-resonant contributions to the currents \cite{Gao2021}.

\section{Conclusion and Outlook}
\label{sec:Conclusion}

We have presented a general approach to handle the observable information encoded in the Bloch states of a crystalline material. The approach is built on projectors to represent the quantum states, which eliminates the gauge dependence. The advantage of the projector formalism is twofold. On the one hand, it offers a natural connection to a geometric description of the complex projective bundle that the Bloch states form. This quantum geometric description allows for a systematic identification of independent wave function properties \cite{Avdoshkin2022, Avdoshkin2024} that serve as minimal building blocks for physical observables. These minimal entities clarify connections between seemingly independent observables and point towards novel phenomena. On the other hand, projector formalism simplifies the theoretical investigation of material properties and their response functions that crucially rely on the complex interplay of internal degrees of freedom, such as spin and atomic orbitals. In particular, topological and geometric insulators and metals that host highly complex quantum states, even in the non-interacting approximation, are included. 

In the first part of this paper, we have introduced the basic concepts of the projector formalism and geometric description for Bloch states, called {\it quantum state geometry}. We closed this section by presenting the essential geometric invariants in the projector formalism involving the quantum metric, Berry curvatures, and their dipoles. We extended the set of geometric invariants beyond the quantum geometric tensor by introducing the corresponding two-state quantities and the quantum geometric connection, which contains novel geometric properties regarding the skewness and torsion tensor. We summarize these quantities in Tab.~\ref{tab:GeometricInvariants}. The presented formalism allows for a systematic construction of novel geometric invariants, whose physical and geometric interpretations remain to be fully understood.

We provide two detailed applications of the formalism: the polarization distribution and the shift and injection current, which was the focus of the companion article \cite{prl2024}. Here, these examples serve as a guiding principle for future applications of formalism to other properties of crystalline materials. In particular, we anticipate that the strategy to identify the relevant geometric invariants for the second-order optical response functions generally simplifies derivations built on standard perturbation-theoretical approaches using Green's function methods, where projectors are naturally introduced via the Bloch Hamiltonian and the spectral representation of the Green's functions \cite{Dupuis2023}. We note that the formalism for momentum-local observables, which is the main focus of this article, can be straightforwardly combined with a perturbative treatment of interaction \cite{Antebi2024}, which requires the inclusion of global geometric invariants and, thus, suggest a rich platform for novel geometric effects.

\begin{acknowledgments}
We thank 
Aris~Alexandradinata,
Ohad~Antebi,
Dan~S.~Borgnia,
Iliya~Esin,
Moritz~M.~Hirschmann, 
Tobias~Holder,
Fernando de Juan,
Wojciech~Jankowski,
Libor~\v Smejkal, 
Michael Sonner, and
Tha\'is~V.~Trevisan
for stimulating discussions. A.~A. was supported by a Kavli ENSI fellowship at UC Berkeley and the National Science Foundation (NSF) Convergence Accelerator Award No. 2235945 at MIT. J.~M. was supported by the German National Academy of Sciences Leopoldina through Grant No. LPDS 2022-06 and, in part, by the Deutsche Forschungsgemeinschaft under Grant cluster of excellence ct.qmat (EXC 2147, Project No. 390858490). J.~E.~M. was supported by the Quantum Materials program under the Director, Office of Science, Office of Basic Energy Sciences, Materials Sciences and Engineering Division, of the U.S. Department of Energy, Contract No. DE-AC02-05CH11231.at Lawrence Berkeley National Laboratory, and a Simons Investigatorship.
\end{acknowledgments}

\appendix

\section{Efficient numerical evaluation of projectors and projector derivatives}
\label{sec:numericalEvaluation}

The numerical derivative requires diagonalizing the Hamiltonian at different momenta, where the obtained Bloch wave functions generically are not expressed in the same gauge. This arbitrary phase highly complicates a stable evaluation of the difference between the Bloch states, which is necessary for a numerical derivative. Explicit gauge fixing procedures are possible but tedious. In contrast, the projectors are gauge invariant by construction, so differences are directly well-defined. The following describes the numerical steps required to obtain the geometric quantities presented in Sec.~\ref{sec:projectors} and in the companion article \cite{prl2024}.

\subsection{Numerical projector construction}

Consider a Bloch Hamiltonian $N_\text{orb}\times N_\text{orb}$ matrix $\hat H(\bk)$ for $N_\text{orb}$ orbitals as a function of momentum $\bk$. When diagonalized for a fixed momentum, we obtain the eigenvalues $E_n(\bk)$ and corresponding orthonormal eigenvectors $|u_n(\bk)\rangle$. For each index $n$, we construct a $N_\text{orb}\times N_\text{orb}$ matrix $\hat P_n(\bk)$ as tensor product between the eigenvector $|u_n\rangle$ and its complex transpose $\langle u_n|$, resulting in $N_\text{orb}$ hermitian matrices satisfying $\hat P_n \hat P_m = \delta_{nm}\hat P_m$ and $\hat H \hat P_n = E_n \hat P_n$ under ordinary matrix multiplication, thus, satisfying Eqs.~\eqref{eqn:basics1} to \eqref{eqn:basics3}. If necessary, projectors onto degenerate or multiple bands are constructed by summing the respective $\hat P_n$. No specific gauge choice is required for $|u_n\rangle$ within this construction as long as $\langle u_n|$ is directly obtained from the corresponding $|u_n\rangle$ by complex transposition. 

\subsection{Numerical derivative construction}

We denote the momentum unit vector in direction $\alpha$ as $\bdir_\alpha$. The first derivative of the projector is obtained by the symmetric finite difference,
\begin{align}
    \partial_\alpha \,\hat P_n(\bk) = \frac{1}{2\lambda}\Big[\hat P_n\big(\bk+\lambda\,\bdir_\alpha\big)-\hat P_n\big(\bk-\lambda\,\bdir_\alpha\big)\Big] + \mathcal{O}(\lambda^{2}) \, ,
\end{align}
with an error of order $\lambda^2$. A complete set of projector derivatives in $d$ spatial dimensions requires $2d$ diagonalizations of the Hamiltonian for each momentum $\bk$. The second derivative is obtained by 
\begin{align}
    \partial_\alpha\partial_\alpha\,\hat P_n(\bk)= \frac{1}{\lambda^2}\Big[&\hat P_n\big(\bk+\lambda\,\bdir_\alpha\big)-2\hat P_n\big(\bk\big)\nonumber\\ &+\hat P_n\big(\bk-\lambda\,\bdir_\alpha\big)\Big] + \mathcal{O}(\lambda^{2}) \, .
\end{align}
A complete set of second-order derivatives requires one further diagonalization of the Hamiltonian. For the off-diagonal second-order derivative, we use
\begin{align}
    \partial_\alpha\partial_\beta\,\hat P_n(\bk)&=\frac{1}{2\lambda^2}\Big[\hat P_n\big(\bk+\lambda\,[\bdir_\alpha+\bdir_\beta]\big)\nonumber\\ &-\hat P_n\big(\bk+\lambda\,\bdir_\alpha\big)-\hat P_n\big(\bk+\lambda\,\bdir_\beta\big)+2\hat P_n\big(\bk\big)\nonumber\\&-\hat P_n\big(\bk-\lambda\,\bdir_\alpha\big)-\hat P_n\big(\bk-\lambda\,\bdir_\beta\big)\nonumber\\&+\hat P_n\big(\bk-\lambda\,[\bdir_\alpha+\bdir_\beta]\big)\Big] + \mathcal{O}(\lambda^{2}) \, ,
\end{align}
which requires $d(d-1)$ further diagonalizations due to the $\bdir_\alpha+\bdir_\beta$ directions. In total, $2d+1+d(d-1)=1+d+d^2$ diagonalizations are required to obtain a complete set of first- and second-order derivatives in $d$ dimensions, that is, $3$ for 1-, $7$ for 2-, and $13$ for 3-dimensions. These numbers can be reduced, for instance, when focusing only on one spatial direction. A $\lambda$ of size $10^{-3}$ to $10^{-5}$ is usually sufficient for stable and reliable numerical results. The numerical accuracy can be checked by various projector identities such as Eqs.~\eqref{eqn:Pzero1} to \eqref{eqn:Pzero3}. 

\section{Strategy to determine closed analytic forms for few-band systems}
\label{sec:Pozo2020}

We describe how to relate the results of Ref.~\onlinecite{Pozo2020} to the presented formalism. Let us consider the generators $\hat M_\alpha$ of SU(N) and expand the Bloch Hamiltonian,
\begin{align}
    \hat H = h_0 \mathds{1}_N + \sum_\alpha h_\alpha \hat M_\alpha
\end{align}
where we have defined $h_0$ related to the trace of the Hamiltonian and $h_\alpha$ as the coefficient of the generator expansion for the traceless part. For a two-band model, such an expansion is given by the well-known form $\hat H = d_0 \mathds{1}_2 + \bd \cdot \boldsymbol{\sigma}$ with Pauli matrices $\boldsymbol{\sigma} = (\sigma_x, \sigma_y, \sigma_z)$. Similarly, we expand the projector onto the band eigenstates as
\begin{align}
    \label{eqn:Pexpansion}
    \hat P_n = \frac{r}{N}\mathds{1}_N+\sum_\alpha P_{n\alpha} \,\hat M_\alpha \,,
\end{align}
where $r$ is the rank of the projector and $P_{n\alpha}$ the expansion coefficient for a given band index $n$ and generator $\alpha$. As an example, the two projectors for a two-band model take the well-known form $\hat P_\pm = \frac{1}{2}\big(\mathds{1}_2 \pm\bn\cdot \boldsymbol{\sigma}\big)$ with $\bn = \bd/|\bd|$. The gauge-invariant matrix elements $M^{nm}_\alpha = \langle u_n|\hat M_\alpha|u_m\rangle$ introduced in Ref.~\onlinecite{Pozo2020} are related to the projector coefficients via
\begin{align}
    M^{nn}_\alpha\equiv \langle u_n|\hat M_\alpha|u_n\rangle = \text{tr}\big[\hat M_\alpha \hat P_n\big] = 2P_{n\alpha}
\end{align}
using $\text{tr}\big[\hat M_\alpha \hat M_\beta\big] = 2\delta_{\alpha\beta}$. Higher-order generators are obtained analogously; for instance, 
\begin{align}
    M^{nm}_\alpha M^{mn}_\beta &\equiv \langle u_n|\hat M_\alpha|u_m\rangle\langle u_m|\hat M_\beta|u_n\rangle \nonumber\\&= \text{tr}\big[\hat P_n \hat M_\alpha \hat P_m \hat M_\beta \big] \nonumber\\&= \sum_{\mu,\nu} P_{n\mu} P_{m\nu} \text{tr}\big[\hat M_\mu\hat M_\alpha \hat M_\nu \hat M_\beta\big] \, ,
\end{align}
where we can evaluate the trace explicitly via the defining equation of the SU(N) algebra, $\hat M_\alpha\hat M_\beta = 2/N\,\delta_{\alpha\beta}\mathds{1}_N + \sum_\gamma S_{\alpha\beta\gamma}\hat M_\gamma$ with complex structure factors $S_{\alpha\beta\gamma} = d_{\alpha\beta\gamma}+i f_{\alpha\beta\gamma}$, capturing the anticommutation and commutation relations as real and imaginary part, respectively \cite{Pozo2020}. We obtain 
\begin{align}
    M^{nm}_\alpha M^{mn}_\beta = \frac{4}{N}P_{n\alpha}P_{m\beta}+2\sum_{\mu,\nu,\gamma} S_{\mu\alpha\gamma}S_{\nu\beta\gamma}P_{n\mu}P_{m\nu}
\end{align}
As long as only gauge-invariant combinations of matrix elements are considered, they can be expressed in terms of the projectors, which we presented as our building block for the geometric description. The geometric objects presented in the main text can be evaluated by inserting Eq.~\eqref{eqn:Pexpansion} and using the SU(N) algebra to evaluate the trace of generators.

\section{Quantum geometric quantities for non-degenerate bands}
\label{sec:GeometryNonDeg}

We present the quantum geometric invariants introduced in Sec.~\ref{sec:ImportantGeometricInvariants} in terms of the Bloch states $|u_n\rangle$ of a non-degenerate band to provide the connections to the existing forms of the geometric invariants in the literature. The band projector of interest reads $\hat P_n = |u_n\rangle\langle u_n|$ in this case. 

\subsection{Quantum geometric tensor}

Expressed in terms of Bloch states, the quantum geometric tensor in Eq.~\eqref{eqn:Qalpha_ab} takes the familiar form
\begin{align}
    Q^n_{\alpha\beta}= \langle \partial_\alpha u_n|\partial_\beta u_n\rangle + \langle u_n|\partial_\alpha u_n\rangle \langle u_n|\partial_\beta u_n \rangle \, ,
\end{align}
from which the quantum metric and Berry curvature expressed in Bloch states is straightforwardly derived. In order to derive the two-state quantum geometric tensor in terms of Bloch states we first obtain 
\begin{align}
    \label{eqn:emn_a}
    \hat e^{mn}_\alpha=i\,\hat P_m\,\partial_\alpha \hat P_n\,\hat P_n=(1-\delta_{mn})\,r^\alpha_{mn}\,|u_m\rangle\langle u_n| \, ,
\end{align}
involving the non-Abelian Berry connection $r^\alpha_{mn}=i\langle u_m|\partial_\alpha u_n\rangle=-i\langle \partial_\alpha u_m|u_n\rangle$. The quantity \eqref{eqn:emn_a} was introduced in Ref.~\onlinecite{Ahn2022} as tangent basis vectors for $U(N)/U(1)^N$ and related to transition dipole matrix elements. Using this identity and focusing on $n\neq m$ for Eq.\eqref{eqn:Qnm} only, we see 
\begin{align}
    &Q^{mn}_{\alpha\beta}=-\text{tr}\big[\hat e^{nm}_\alpha\hat e^{mn}_\beta\big]\\&=-(1-\delta_{nm})^2\,r^\alpha_{nm}r^\beta_{mn}\,\text{tr}\Big[|u_n\rangle \langle u_m|u_m\rangle \langle u_n|\Big]\\&=-(1-\delta_{nm})\,r^\alpha_{nm}r^\beta_{mn} \, .
    \label{eqn:QmnExplicit}
\end{align}
which establishes the connection between the product of the non-Abelian Berry connections and the two-state quantum geometric tensor.

\subsection{Quantum geometric connection}
\label{app:quantumgeometricconnection}

Inserting the explicit form of band projector onto a non-degenerate band into Eq.~\eqref{eqn:Qalphabetagamma}, we obtain 
\begin{align}
    &Q^n_{\alpha;\beta\gamma}
    =\frac{1}{2}\langle\partial_\alpha u_n|\partial_\beta\partial_\gamma u_n\rangle + \frac{1}{2}\langle u_n|\partial_\alpha u_n\rangle\langle u_n|\partial_\beta\partial_\gamma u_n\rangle \nonumber \\[1mm] &-\langle \partial_\alpha u_n|\partial_\beta u_n\rangle\langle u_n|\partial_\gamma u_n \rangle \nonumber\\[2mm]&- \langle u_n|\partial_\alpha u_n\rangle\langle u_n | \partial_\beta u_n\rangle \langle u_n | \partial_\gamma u_n\rangle + (\beta\leftrightarrow \gamma) \, .
\end{align}
The expression shows explicitly that the projector form yields a much compacter form even for non-degenerate bands. In order to derive the explicit form of the two-state quantum geometric connection given in Eq.~\eqref{eqn:Cmn_abg} for $n\neq m$, we use Eq.~\eqref{eqn:emn_a} and obtain
%
\begin{align}
    &C^{mn}_{\alpha;\beta\gamma}=-\text{tr}\big[\hat e^{nm}_\beta\,\partial^{}_\alpha\hat e^{mn}_\gamma\big]\\&=-(1-\delta_{nm})^2 r^\beta_{nm}\text{tr}\Big[|u_n\rangle \langle u_m|\,\partial_\alpha\big(r^\gamma_{mn} |u_m\rangle \langle u_n|\big)\!\Big]
    \\&=-(1-\delta_{nm})\,r^\beta_{nm}\,\Big(\partial_\alpha\,r^\gamma_{mn}-i\big(\xi^\alpha_m-\xi^\alpha_n\big)r^\gamma_{mn}\Big)\\&=-(1-\delta_{nm})\,r^\beta_{nm}\,r^\gamma_{mn;\alpha}
    \label{eqn:CmnExplicit}
\end{align}
%
with band Berry connection $\xi^\alpha_n=r^\alpha_{nn}=i\langle u_n|\partial_\alpha u_n\rangle$ and covariant derivative of the non-Abelian Berry connection, $r^\gamma_{mn;a}=\partial_\alpha r^\gamma_{mn}-i(\xi^\alpha_m-\xi^\alpha_n)r^\gamma_{mn}$.

\subsection{Interband Wilson loop and shift vector} \label{sec:AppendixWilsonShift}
We derive the explicit expression for the interband Wilson loop in Eq.~\eqref{eqn:interBandWilson} for non-degenerate bands using Eq.~\eqref{eqn:emn_a}. We obtain 
\begin{align}
    &W^{mn}_{\alpha\beta}(\bk,\bq) = \text{tr}\big[\hat e^{nm}_\beta(\bk)\,\hat e^{mn}_\alpha(\bk+\bq)\big] \\ &\,\,\,\,\,\,=(1-\delta_{nm})\,r_{nm}^\beta(\bk)\,r_{mn}^\alpha(\bk+\bq)\nonumber\\&\,\,\,\,\,\,\,\,\,\,\times\text{tr}\big[|u_n(\bk)\rangle\langle u_n(\bk)|u_m(\bk+\bq)\rangle\langle u_m(\bk+\bq)|\big]\\ &\,\,\,\,\,\,= (1-\delta_{nm})\,r^\alpha_{mn}(\bk+\bq)\, r^\beta_{nm}(\bk) \nonumber \\ &\,\,\,\,\,\,\,\,\,\,\times \langle u_n(\bk)|u_m(\bk+\bq)\rangle  \langle u_n(\bk+\bq)|u_n(\bk)\rangle \, ,
\end{align}
where we recover the expression given in Ref.~\cite{Wang2022} with explicitly subtracted intraband contribution $n=m$. Using the definition of the shift vector in Eq.~\eqref{eqn:shiftvector} and inserting the expressions for the two-state quantum geometric tensor and quantum geometric connection given in Eqs.~\eqref{eqn:QmnExplicit} and \eqref{eqn:CmnExplicit} gives
\begin{align}
    R^{mn}_{\alpha\beta} &= i\frac{C^{mn}_{\alpha;\beta\beta}}{Q^{mn}_{\beta\beta}} = i\frac{r^\beta_{nm}r^\beta_{mn;\alpha}}{r^\beta_{nm}r^\beta_{mn}} = i\frac{r^\beta_{mn;\alpha}}{r^\beta_{mn}} \\&=i\,\frac{\partial_\alpha r^\beta_{mn}-i(\xi^\alpha_m-\xi^\alpha_n)r^\beta_{mn}}{r^\beta_{mn}} \\
    &=\xi^\alpha_m-\xi^\alpha_n+i\, \partial_\alpha \ln\,r^\beta_{mn} \, .
\end{align}
We recover the standard definition of the shift current \cite{Sipe2000, Ahn2022} in terms of the band Berry connection $\xi^\alpha_n=r^\alpha_{nn}=i\langle u_n|\partial_\alpha u_n\rangle$ and the non-Abelian Berry connection $r^\alpha_{mn}=i\langle u_m|\partial_\alpha u_n\rangle$.

\section{Derivation of the generating function of the polarization distribution}
\label{sec:PolarizationAppendix}

We provide the derivation from Eq.~\eqref{eq:C_of_q} to \eqref{eq:C_q_expr}, which follows the steps presented first in Ref.~\onlinecite{Avdoshkin2022} by one of us. We note the identity
\begin{align} \label{eq:c_sum}
    &C(\bq) + C(\bq') \nonumber\\&= \int_\text{BZ} \log \det \Big(\langle u_i(\bk)|\,\hat P_{\bk+\bq}\,|u_j(\bk+\bq + \bq')\rangle \Big)_{ij} \, ,    
\end{align}
where $\hat P_\bk = \sum_{n\in\text{occ}} |u_n(\bk)\rangle \langle u_n(\bk)|$ is the projector onto all occupied bands. Using this, we obtain
\begin{align}
    &\log C(-\bq - \bq') + \log C(\bq) + \log C(\bq') \nonumber\\&= \int_\text{BZ} \log \text{det}_{u_\bk}\! \big[\hat P_\bk\, \hat P_{\bk+\bq}\, \hat P_{\bk+\bq + \bq'}\, \hat P_\bk\big] \, .
\end{align}
We note that
\begin{align}
    \log \!\Big[C(\bq) C(-\bq)\Big] = \int_\text{BZ}\log \text{det}_{u_\bk}\!\big[\hat P_\bk\, \hat P_{\bk+\bq}\,\hat P_\bk \big] \,,
\end{align}
so that we arrive at
\begin{align}
    &\log \frac{C(\bq + \bq')}{C(\bq) C(\bq')} =-\int_\text{BZ} \log \text{det}_{u_\bk}\! \big[\hat P_\bk \,\hat P_{\bk+\bq}\, \hat P_{\bk+\bq + \bq'}\, \hat P_\bk\big] \nonumber\\&\hspace{25mm}+ \int_\text{BZ} \log \text{det}_{u_\bk}\!\big[P_\bk \, \hat P_{\bk+\bq + \bq'}\, \hat P_\bk\big] \, .
\end{align}
It is convenient to define $b_{\bv}(t) = \log C(\bv t)$, for which we obtain
\begin{align} \label{eq:dif_eqn}
    &\partial_t b_{\bv}(t) - \partial_t b_{\bv}(0) \nonumber\\& =-\int_\text{BZ} \text{tr}\Big[\big(\hat P_\bk\, \hat P_{\bk + \bv t}\, \hat P_\bk\big)^{-1} \, \hat P_\bk \,\hat P_{\bk + \bv t}\,\big(\partial_{t}\hat P_{\bk + \bv t}\big)\,\hat P_\bk \big] \nonumber\\& \hspace{4mm}+ \int_\text{BZ} \text{tr}\Big[ \big(\hat P_\bk\, \hat P_{\bk + \bv t}\, \hat P_\bk\big)^{-1} \,\hat P_\bk \,\big(\partial_t \hat P_{\bk + \bv t}\big)\,\hat P_\bk \Big] \, 
\end{align}
via the Jacobi's formula
\begin{align}
    \frac{d}{d t}\det A(t) = \det\!\big[A(t)\big]\,\text{tr}\bigg[A^{-1}(t)\frac{d}{d t}A(t)\bigg].
\end{align}
Both terms in Eq.~\eqref{eq:dif_eqn} can be combined via the projector identity $\partial_t \hat P_t = (\partial_t \hat P_t)\hat P_t + \hat P_t (\partial_t \hat P_t)$, such that we identify 
\begin{align}
    \label{eqn:Aabk}
    \mathcal{A}_{\alpha}^{\bk}(\bk') = \text{tr}\Big[\hat P_{\bk}\,\big(\hat P_{\bk} \, \hat P_{\bk'}\,\hat P_{\bk}\big)^{-1} \,\hat P_{\bk}\,\big(\partial_{\alpha}\,\hat P_{\bk'}\big) \,\hat P_{\bk'} \Big] \, .
\end{align}
With that, we write
\begin{align}
    \partial_t b_{\bv}(t) - \partial_t b_{\bv}(0) = \sum_\alpha \n^{}_\alpha \,\mathcal{A}_\alpha^{\bk}(\bk + \bv t) \, ,
\end{align}
where $\n_\alpha$ is the unit vector in the direction of $\bv$. Integrating this equation, we arrive at Eq.~\eqref{eq:C_q_expr}, where the constant $\mathcal{A}_\alpha = \sum_{n\in\text{occ}}\int_\text{BZ} \mathcal{A}^n_\alpha(\bk)$ is fixed by the known relation between the mean polarization and the Berry connection.

\section{Derivation of the two remaining contributions to the injection and shift current}
\label{sec:shiftcurrentAppendix}

\subsection{Vanishing of the contribution $\sigma^{\aCord;\bCord\cCord}_{(0)}$}

We show that $\sigma^{\aCord;\bCord\cCord}_{(0)}$ defined as 
\begin{align}
    \label{eqn:sigma0}
    \sigma^{\aCord;\bCord\cCord}_{(0)} &= \text{tr}\Bigg[ \hat P_\text{occ} \bigg(\!\partial_\aCord\partial_\bCord\partial_\cCord \hat H \!+ \!\bigg[\partial_\bCord\partial_\cCord \hat H\,,\, \frac{\partial_\aCord \hat H}{- \epsilon + i \gamma}\bigg] \bigg)\!\Bigg],
\end{align}
vanishes in leading orders of $\gamma\ll|\epsilon_{nm}|$. Using the separation of the fraction into two contribution as given in Eq.~\eqref{eqn:gamma_expansionMain}, we see that inserting of the first term leads to a vanishing contribution via Eq.~\eqref{eqn:Pocc_identity1Main},
\begin{align}
     -\frac{i}{\gamma}\sum_{n}(\partial_\aCord E_n)\,\text{tr}\bigg[ \hat P_\text{occ} \Big[\partial_\bCord\partial_\cCord \hat H\,,\, \hat P_n\Big]\bigg] = 0 \, .
\end{align}
Inserting the second term to leading order in $\gamma\ll|\epsilon_{nm}|$ reads 
\begin{align}
    &\sum_n\text{tr}\Big[ \hat P_\text{occ} \big[\partial_\bCord\partial_\cCord \hat H\,,\,(\partial_\aCord \hat P_n)\,\hat P_n\big] \Big] \, .
\end{align}   
Inserting the explicit form of $\hat P_\text{occ}$ given in Eq.~\eqref{eqn:PoccFermi} and using the derivative identity $\hat P_n (\partial_\aCord \hat P_m) = \delta_{nm}\partial_\aCord \hat P_n - (\partial_\aCord \hat P_n)\hat P_m$ leads to
\begin{align}
    &\sum_{n,m}f_m\,\text{tr}\Big[(\partial_\aCord \hat P_n)\hat P_n\hat P_m (\partial_\bCord\partial_\cCord \hat H)- \hat P_m (\partial_\aCord \hat P_n)\hat P_n (\partial_\bCord\partial_\cCord \hat H) \Big] \nonumber \\
    &= \sum_{n,m}f_m\,\text{tr}\Big[ (\partial_\aCord \hat P_m)\hat P_n (\partial_\bCord\partial_\cCord \hat H) \Big] \\
    &= \sum_{n}f_n\,\text{tr}\Big[ (\partial_\aCord \hat P_n) (\partial_\bCord\partial_\cCord \hat H) \Big] \, .
\end{align}
We performed the sum $\sum_n \hat P_n = \hat 1$ in the last step. We see that the result is equal to the $-\text{tr}\big[\hat P_\text{occ} \,(\partial_\aCord\partial_\bCord\partial_\cCord \hat H)\big]$ under the momentum integral assuming a momentum-constant $f_n$ and performing a partial integration in the $\aCord$-direction, so that $\sigma^{abc}_{(0)}$ vanishes.  

\subsection{Leading-order contributions to $\sigma^{\aCord;\bCord\cCord}_{(1)}(\omega)$}

We evaluate the remaining contribution to Eq.~\eqref{eqn:sigmaTensor}, which takes the form $\sigma^{\aCord;\bCord\cCord}_{(1)}(\omega) = A^{\bCord;\aCord\cCord}(\omega)+A^{\cCord;\aCord\bCord}(-\omega)$ with
\begin{align}
    &A^{\aCord;\bCord\cCord}(\omega) \equiv \text{tr}\Bigg[\hat P_\text{occ}\,\bigg[\frac{\partial_\aCord \hat H}{\omega+\epsilon+i\Gamma},\partial_\bCord\partial_\cCord \hat H\bigg]\Bigg] \, .
\end{align}
We decompose the fraction via Eq.~\eqref{eqn:Gamma_expansionMain}. The first term leads to
\begin{align}
   \text{tr}\Big[\hat P_\text{occ}\,\big[\hat P_n,\partial_\bCord\partial_\cCord \hat H\big]\Big] = 0 \, ,
\end{align}
which vanishes due to identity \eqref{eqn:Pocc_identity1Main}. The second term involves
\begin{align}
     &\text{tr}\bigg[\hat P_\text{occ}\,\Big[\hat P_n(\partial_\aCord \hat P_m)\hat P_m\,,\,\partial_\bCord\partial_\cCord \hat H\Big]\bigg] \\ &= f_{nm}\text{tr}\big[(\partial_\bCord\partial_\cCord \hat H)\hat P_n(\partial_\aCord\hat P_m)\hat P_m\big] \, ,
\end{align}
which we simplified via identity \eqref{eqn:Pocc_identity2Main}. Determining the prefactors and combining the contributions leads to the form given in Eq.~\eqref{eqn:sigma1_final_main}.

\bibliography{arxiv_PRB_v2}

\begin{thebibliography}{52}%
\makeatletter
\providecommand \@ifxundefined [1]{%
 \@ifx{#1\undefined}
}%
\providecommand \@ifnum [1]{%
 \ifnum #1\expandafter \@firstoftwo
 \else \expandafter \@secondoftwo
 \fi
}%
\providecommand \@ifx [1]{%
 \ifx #1\expandafter \@firstoftwo
 \else \expandafter \@secondoftwo
 \fi
}%
\providecommand \natexlab [1]{#1}%
\providecommand \enquote  [1]{``#1''}%
\providecommand \bibnamefont  [1]{#1}%
\providecommand \bibfnamefont [1]{#1}%
\providecommand \citenamefont [1]{#1}%
\providecommand \href@noop [0]{\@secondoftwo}%
\providecommand \href [0]{\begingroup \@sanitize@url \@href}%
\providecommand \@href[1]{\@@startlink{#1}\@@href}%
\providecommand \@@href[1]{\endgroup#1\@@endlink}%
\providecommand \@sanitize@url [0]{\catcode `\\12\catcode `\$12\catcode `\&12\catcode `\#12\catcode `\^12\catcode `\_12\catcode `\%12\relax}%
\providecommand \@@startlink[1]{}%
\providecommand \@@endlink[0]{}%
\providecommand \url  [0]{\begingroup\@sanitize@url \@url }%
\providecommand \@url [1]{\endgroup\@href {#1}{\urlprefix }}%
\providecommand \urlprefix  [0]{URL }%
\providecommand \Eprint [0]{\href }%
\providecommand \doibase [0]{https://doi.org/}%
\providecommand \selectlanguage [0]{\@gobble}%
\providecommand \bibinfo  [0]{\@secondoftwo}%
\providecommand \bibfield  [0]{\@secondoftwo}%
\providecommand \translation [1]{[#1]}%
\providecommand \BibitemOpen [0]{}%
\providecommand \bibitemStop [0]{}%
\providecommand \bibitemNoStop [0]{.\EOS\space}%
\providecommand \EOS [0]{\spacefactor3000\relax}%
\providecommand \BibitemShut  [1]{\csname bibitem#1\endcsname}%
\let\auto@bib@innerbib\@empty
\bibitem [{\citenamefont {Thouless}\ \emph {et~al.}(1982)\citenamefont {Thouless}, \citenamefont {Kohmoto}, \citenamefont {Nightingale},\ and\ \citenamefont {den Nijs}}]{Thouless1982}%
  \BibitemOpen
  \bibfield  {author} {\bibinfo {author} {\bibfnamefont {D.~J.}\ \bibnamefont {Thouless}}, \bibinfo {author} {\bibfnamefont {M.}~\bibnamefont {Kohmoto}}, \bibinfo {author} {\bibfnamefont {M.~P.}\ \bibnamefont {Nightingale}},\ and\ \bibinfo {author} {\bibfnamefont {M.}~\bibnamefont {den Nijs}},\ }\bibfield  {title} {\bibinfo {title} {{Quantized Hall Conductance in a Two-Dimensional Periodic Potential}},\ }\href {https://doi.org/10.1103/PhysRevLett.49.405} {\bibfield  {journal} {\bibinfo  {journal} {Phys. Rev. Lett.}\ }\textbf {\bibinfo {volume} {49}},\ \bibinfo {pages} {405} (\bibinfo {year} {1982})}\BibitemShut {NoStop}%
\bibitem [{\citenamefont {Chang}\ and\ \citenamefont {Niu}(1995)}]{Chang1995}%
  \BibitemOpen
  \bibfield  {author} {\bibinfo {author} {\bibfnamefont {M.-C.}\ \bibnamefont {Chang}}\ and\ \bibinfo {author} {\bibfnamefont {Q.}~\bibnamefont {Niu}},\ }\bibfield  {title} {\bibinfo {title} {{Berry Phase, Hyperorbits, and the Hofstadter Spectrum}},\ }\href {https://doi.org/10.1103/PhysRevLett.75.1348} {\bibfield  {journal} {\bibinfo  {journal} {Phys. Rev. Lett.}\ }\textbf {\bibinfo {volume} {75}},\ \bibinfo {pages} {1348} (\bibinfo {year} {1995})}\BibitemShut {NoStop}%
\bibitem [{\citenamefont {Chang}\ and\ \citenamefont {Niu}(1996)}]{Chang1996}%
  \BibitemOpen
  \bibfield  {author} {\bibinfo {author} {\bibfnamefont {M.-C.}\ \bibnamefont {Chang}}\ and\ \bibinfo {author} {\bibfnamefont {Q.}~\bibnamefont {Niu}},\ }\bibfield  {title} {\bibinfo {title} {{Berry phase, hyperorbits, and the Hofstadter spectrum: Semiclassical dynamics in magnetic Bloch bands}},\ }\href {https://doi.org/10.1103/PhysRevB.53.7010} {\bibfield  {journal} {\bibinfo  {journal} {Phys. Rev. B}\ }\textbf {\bibinfo {volume} {53}},\ \bibinfo {pages} {7010} (\bibinfo {year} {1996})}\BibitemShut {NoStop}%
\bibitem [{\citenamefont {Sundaram}\ and\ \citenamefont {Niu}(1999)}]{Sundaram1999}%
  \BibitemOpen
  \bibfield  {author} {\bibinfo {author} {\bibfnamefont {G.}~\bibnamefont {Sundaram}}\ and\ \bibinfo {author} {\bibfnamefont {Q.}~\bibnamefont {Niu}},\ }\bibfield  {title} {\bibinfo {title} {{Wave-packet dynamics in slowly perturbed crystals: Gradient corrections and Berry-phase effects}},\ }\href {https://doi.org/10.1103/PhysRevB.59.14915} {\bibfield  {journal} {\bibinfo  {journal} {Phys. Rev. B}\ }\textbf {\bibinfo {volume} {59}},\ \bibinfo {pages} {14915} (\bibinfo {year} {1999})}\BibitemShut {NoStop}%
\bibitem [{\citenamefont {Avron}\ \emph {et~al.}(1983)\citenamefont {Avron}, \citenamefont {Seiler},\ and\ \citenamefont {Simon}}]{Avron1983}%
  \BibitemOpen
  \bibfield  {author} {\bibinfo {author} {\bibfnamefont {J.~E.}\ \bibnamefont {Avron}}, \bibinfo {author} {\bibfnamefont {R.}~\bibnamefont {Seiler}},\ and\ \bibinfo {author} {\bibfnamefont {B.}~\bibnamefont {Simon}},\ }\bibfield  {title} {\bibinfo {title} {Homotopy and quantization in condensed matter physics},\ }\href {https://doi.org/10.1103/PhysRevLett.51.51} {\bibfield  {journal} {\bibinfo  {journal} {Phys. Rev. Lett.}\ }\textbf {\bibinfo {volume} {51}},\ \bibinfo {pages} {51} (\bibinfo {year} {1983})}\BibitemShut {NoStop}%
\bibitem [{\citenamefont {Provost}(1980)}]{Provost1980}%
  \BibitemOpen
  \bibfield  {author} {\bibinfo {author} {\bibfnamefont {G.}~\bibnamefont {Provost}, \bibfnamefont {J.~P.~Vallee}},\ }\bibfield  {title} {\bibinfo {title} {{Riemannian structure on manifolds of quantum states}},\ }\href {https://doi.org/10.1007/BF02193559} {\bibfield  {journal} {\bibinfo  {journal} {Communications in Mathematical Physics}\ }\textbf {\bibinfo {volume} {76}} (\bibinfo {year} {1980})}\BibitemShut {NoStop}%
\bibitem [{\citenamefont {Avdoshkin}\ \emph {et~al.}(2024)\citenamefont {Avdoshkin}, \citenamefont {Mitscherling},\ and\ \citenamefont {Moore}}]{prl2024}%
  \BibitemOpen
  \bibfield  {author} {\bibinfo {author} {\bibfnamefont {A.}~\bibnamefont {Avdoshkin}}, \bibinfo {author} {\bibfnamefont {J.}~\bibnamefont {Mitscherling}},\ and\ \bibinfo {author} {\bibfnamefont {J.~E.}\ \bibnamefont {Moore}},\ }\href {https://arxiv.org/abs/2409.16358} {\bibinfo {title} {The multi-state geometry of shift current and polarization}} (\bibinfo {year} {2024}),\ \Eprint {https://arxiv.org/abs/2409.16358} {arXiv:2409.16358 [cond-mat.str-el]} \BibitemShut {NoStop}%
\bibitem [{\citenamefont {Pozo}\ and\ \citenamefont {de~Juan}(2020)}]{Pozo2020}%
  \BibitemOpen
  \bibfield  {author} {\bibinfo {author} {\bibfnamefont {O.}~\bibnamefont {Pozo}}\ and\ \bibinfo {author} {\bibfnamefont {F.}~\bibnamefont {de~Juan}},\ }\bibfield  {title} {\bibinfo {title} {Computing observables without eigenstates: Applications to bloch hamiltonians},\ }\href {https://doi.org/10.1103/PhysRevB.102.115138} {\bibfield  {journal} {\bibinfo  {journal} {Phys. Rev. B}\ }\textbf {\bibinfo {volume} {102}},\ \bibinfo {pages} {115138} (\bibinfo {year} {2020})}\BibitemShut {NoStop}%
\bibitem [{\citenamefont {Graf}\ and\ \citenamefont {Pi\'echon}(2021)}]{Graf2021}%
  \BibitemOpen
  \bibfield  {author} {\bibinfo {author} {\bibfnamefont {A.}~\bibnamefont {Graf}}\ and\ \bibinfo {author} {\bibfnamefont {F.}~\bibnamefont {Pi\'echon}},\ }\bibfield  {title} {\bibinfo {title} {{Berry curvature and quantum metric in $N$-band systems: An eigenprojector approach}},\ }\href {https://doi.org/10.1103/PhysRevB.104.085114} {\bibfield  {journal} {\bibinfo  {journal} {Phys. Rev. B}\ }\textbf {\bibinfo {volume} {104}},\ \bibinfo {pages} {085114} (\bibinfo {year} {2021})}\BibitemShut {NoStop}%
\bibitem [{\citenamefont {Mera}\ and\ \citenamefont {Ozawa}(2021)}]{Mera2021}%
  \BibitemOpen
  \bibfield  {author} {\bibinfo {author} {\bibfnamefont {B.}~\bibnamefont {Mera}}\ and\ \bibinfo {author} {\bibfnamefont {T.}~\bibnamefont {Ozawa}},\ }\bibfield  {title} {\bibinfo {title} {{K\"ahler geometry and Chern insulators: Relations between topology and the quantum metric}},\ }\href {https://doi.org/10.1103/PhysRevB.104.045104} {\bibfield  {journal} {\bibinfo  {journal} {Phys. Rev. B}\ }\textbf {\bibinfo {volume} {104}},\ \bibinfo {pages} {045104} (\bibinfo {year} {2021})}\BibitemShut {NoStop}%
\bibitem [{\citenamefont {Mera}\ and\ \citenamefont {Mitscherling}(2022)}]{Mera2022}%
  \BibitemOpen
  \bibfield  {author} {\bibinfo {author} {\bibfnamefont {B.}~\bibnamefont {Mera}}\ and\ \bibinfo {author} {\bibfnamefont {J.}~\bibnamefont {Mitscherling}},\ }\bibfield  {title} {\bibinfo {title} {Nontrivial quantum geometry of degenerate flat bands},\ }\href {https://doi.org/10.1103/PhysRevB.106.165133} {\bibfield  {journal} {\bibinfo  {journal} {Phys. Rev. B}\ }\textbf {\bibinfo {volume} {106}},\ \bibinfo {pages} {165133} (\bibinfo {year} {2022})}\BibitemShut {NoStop}%
\bibitem [{\citenamefont {Avdoshkin}\ and\ \citenamefont {Popov}(2023)}]{Avdoshkin2022}%
  \BibitemOpen
  \bibfield  {author} {\bibinfo {author} {\bibfnamefont {A.}~\bibnamefont {Avdoshkin}}\ and\ \bibinfo {author} {\bibfnamefont {F.~K.}\ \bibnamefont {Popov}},\ }\bibfield  {title} {\bibinfo {title} {Extrinsic geometry of quantum states},\ }\href {https://doi.org/10.1103/PhysRevB.107.245136} {\bibfield  {journal} {\bibinfo  {journal} {Phys. Rev. B}\ }\textbf {\bibinfo {volume} {107}},\ \bibinfo {pages} {245136} (\bibinfo {year} {2023})}\BibitemShut {NoStop}%
\bibitem [{\citenamefont {Avdoshkin}(2024)}]{Avdoshkin2024}%
  \BibitemOpen
  \bibfield  {author} {\bibinfo {author} {\bibfnamefont {A.}~\bibnamefont {Avdoshkin}},\ }\bibfield  {title} {\bibinfo {title} {Geometry of degenerate quantum states, configurations of $ m $-planes and invariants on complex grassmannians},\ }\href@noop {} {\bibfield  {journal} {\bibinfo  {journal} {arXiv preprint arXiv:2404.03234}\ } (\bibinfo {year} {2024})}\BibitemShut {NoStop}%
\bibitem [{\citenamefont {Antebi}\ \emph {et~al.}(2024)\citenamefont {Antebi}, \citenamefont {Mitscherling},\ and\ \citenamefont {Holder}}]{Antebi2024}%
  \BibitemOpen
  \bibfield  {author} {\bibinfo {author} {\bibfnamefont {O.}~\bibnamefont {Antebi}}, \bibinfo {author} {\bibfnamefont {J.}~\bibnamefont {Mitscherling}},\ and\ \bibinfo {author} {\bibfnamefont {T.}~\bibnamefont {Holder}},\ }\href@noop {} {\bibinfo {title} {The drude weight of a flatband metal}} (\bibinfo {year} {2024}),\ \Eprint {https://arxiv.org/abs/2407.09599} {arXiv:2407.09599 [cond-mat.str-el]} \BibitemShut {NoStop}%
\bibitem [{\citenamefont {Dupuis}(2023)}]{Dupuis2023}%
  \BibitemOpen
  \bibfield  {author} {\bibinfo {author} {\bibfnamefont {N.}~\bibnamefont {Dupuis}},\ }\href {https://doi.org/10.1142/q0409} {\emph {\bibinfo {title} {Field Theory of Condensed Matter and Ultracold Gases}}}\ (\bibinfo  {publisher} {WORLD SCIENTIFIC (EUROPE)},\ \bibinfo {year} {2023})\ \Eprint {https://arxiv.org/abs/https://www.worldscientific.com/doi/pdf/10.1142/q0409} {https://www.worldscientific.com/doi/pdf/10.1142/q0409} \BibitemShut {NoStop}%
\bibitem [{\citenamefont {Topp}\ \emph {et~al.}(2019)\citenamefont {Topp}, \citenamefont {Jotzu}, \citenamefont {McIver}, \citenamefont {Xian}, \citenamefont {Rubio},\ and\ \citenamefont {Sentef}}]{Topp2019}%
  \BibitemOpen
  \bibfield  {author} {\bibinfo {author} {\bibfnamefont {G.~E.}\ \bibnamefont {Topp}}, \bibinfo {author} {\bibfnamefont {G.}~\bibnamefont {Jotzu}}, \bibinfo {author} {\bibfnamefont {J.~W.}\ \bibnamefont {McIver}}, \bibinfo {author} {\bibfnamefont {L.}~\bibnamefont {Xian}}, \bibinfo {author} {\bibfnamefont {A.}~\bibnamefont {Rubio}},\ and\ \bibinfo {author} {\bibfnamefont {M.~A.}\ \bibnamefont {Sentef}},\ }\bibfield  {title} {\bibinfo {title} {{Topological Floquet engineering of twisted bilayer graphene}},\ }\href {https://doi.org/10.1103/PhysRevResearch.1.023031} {\bibfield  {journal} {\bibinfo  {journal} {Phys. Rev. Res.}\ }\textbf {\bibinfo {volume} {1}},\ \bibinfo {pages} {023031} (\bibinfo {year} {2019})}\BibitemShut {NoStop}%
\bibitem [{\citenamefont {Topp}\ \emph {et~al.}(2021)\citenamefont {Topp}, \citenamefont {Eckhardt}, \citenamefont {Kennes}, \citenamefont {Sentef},\ and\ \citenamefont {T\"orm\"a}}]{Topp2021}%
  \BibitemOpen
  \bibfield  {author} {\bibinfo {author} {\bibfnamefont {G.~E.}\ \bibnamefont {Topp}}, \bibinfo {author} {\bibfnamefont {C.~J.}\ \bibnamefont {Eckhardt}}, \bibinfo {author} {\bibfnamefont {D.~M.}\ \bibnamefont {Kennes}}, \bibinfo {author} {\bibfnamefont {M.~A.}\ \bibnamefont {Sentef}},\ and\ \bibinfo {author} {\bibfnamefont {P.}~\bibnamefont {T\"orm\"a}},\ }\bibfield  {title} {\bibinfo {title} {Light-matter coupling and quantum geometry in moir\'e materials},\ }\href {https://doi.org/10.1103/PhysRevB.104.064306} {\bibfield  {journal} {\bibinfo  {journal} {Phys. Rev. B}\ }\textbf {\bibinfo {volume} {104}},\ \bibinfo {pages} {064306} (\bibinfo {year} {2021})}\BibitemShut {NoStop}%
\bibitem [{\citenamefont {Tai}\ and\ \citenamefont {Claassen}(2023)}]{Tai2023}%
  \BibitemOpen
  \bibfield  {author} {\bibinfo {author} {\bibfnamefont {W.~T.}\ \bibnamefont {Tai}}\ and\ \bibinfo {author} {\bibfnamefont {M.}~\bibnamefont {Claassen}},\ }\href {https://arxiv.org/abs/2303.01597} {\bibinfo {title} {Quantum-geometric light-matter coupling in correlated quantum materials}} (\bibinfo {year} {2023}),\ \Eprint {https://arxiv.org/abs/2303.01597} {arXiv:2303.01597 [cond-mat.str-el]} \BibitemShut {NoStop}%
\bibitem [{\citenamefont {Chen}\ and\ \citenamefont {von Gersdorff}(2022)}]{Chen2022}%
  \BibitemOpen
  \bibfield  {author} {\bibinfo {author} {\bibfnamefont {W.}~\bibnamefont {Chen}}\ and\ \bibinfo {author} {\bibfnamefont {G.}~\bibnamefont {von Gersdorff}},\ }\bibfield  {title} {\bibinfo {title} {{Measurement of interaction-dressed Berry curvature and quantum metric in solids by optical absorption}},\ }\href {https://doi.org/10.21468/SciPostPhysCore.5.3.040} {\bibfield  {journal} {\bibinfo  {journal} {SciPost Phys. Core}\ }\textbf {\bibinfo {volume} {5}},\ \bibinfo {pages} {040} (\bibinfo {year} {2022})}\BibitemShut {NoStop}%
\bibitem [{\citenamefont {Kashihara}\ \emph {et~al.}(2023)\citenamefont {Kashihara}, \citenamefont {Michishita},\ and\ \citenamefont {Peters}}]{Kashihara2023}%
  \BibitemOpen
  \bibfield  {author} {\bibinfo {author} {\bibfnamefont {T.}~\bibnamefont {Kashihara}}, \bibinfo {author} {\bibfnamefont {Y.}~\bibnamefont {Michishita}},\ and\ \bibinfo {author} {\bibfnamefont {R.}~\bibnamefont {Peters}},\ }\bibfield  {title} {\bibinfo {title} {{Quantum metric on the Brillouin zone in correlated electron systems and its relation to topology for Chern insulators}},\ }\href {https://doi.org/10.1103/PhysRevB.107.125116} {\bibfield  {journal} {\bibinfo  {journal} {Phys. Rev. B}\ }\textbf {\bibinfo {volume} {107}},\ \bibinfo {pages} {125116} (\bibinfo {year} {2023})}\BibitemShut {NoStop}%
\bibitem [{\citenamefont {Verma}\ and\ \citenamefont {Queiroz}(2024)}]{Verma2024}%
  \BibitemOpen
  \bibfield  {author} {\bibinfo {author} {\bibfnamefont {N.}~\bibnamefont {Verma}}\ and\ \bibinfo {author} {\bibfnamefont {R.}~\bibnamefont {Queiroz}},\ }\href@noop {} {\bibinfo {title} {Instantaneous response and quantum geometry of insulators}} (\bibinfo {year} {2024}),\ \Eprint {https://arxiv.org/abs/2403.07052} {arXiv:2403.07052 [cond-mat.mes-hall]} \BibitemShut {NoStop}%
\bibitem [{\citenamefont {Ahn}\ \emph {et~al.}(2020)\citenamefont {Ahn}, \citenamefont {Guo},\ and\ \citenamefont {Nagaosa}}]{Ahn2020}%
  \BibitemOpen
  \bibfield  {author} {\bibinfo {author} {\bibfnamefont {J.}~\bibnamefont {Ahn}}, \bibinfo {author} {\bibfnamefont {G.-Y.}\ \bibnamefont {Guo}},\ and\ \bibinfo {author} {\bibfnamefont {N.}~\bibnamefont {Nagaosa}},\ }\bibfield  {title} {\bibinfo {title} {Low-frequency divergence and quantum geometry of the bulk photovoltaic effect in topological semimetals},\ }\href {https://doi.org/10.1103/PhysRevX.10.041041} {\bibfield  {journal} {\bibinfo  {journal} {Phys. Rev. X}\ }\textbf {\bibinfo {volume} {10}},\ \bibinfo {pages} {041041} (\bibinfo {year} {2020})}\BibitemShut {NoStop}%
\bibitem [{\citenamefont {Ahn}\ \emph {et~al.}(2022)\citenamefont {Ahn}, \citenamefont {Guo}, \citenamefont {Nagaosa},\ and\ \citenamefont {Vishwanath}}]{Ahn2022}%
  \BibitemOpen
  \bibfield  {author} {\bibinfo {author} {\bibfnamefont {J.}~\bibnamefont {Ahn}}, \bibinfo {author} {\bibfnamefont {G.-Y.}\ \bibnamefont {Guo}}, \bibinfo {author} {\bibfnamefont {N.}~\bibnamefont {Nagaosa}},\ and\ \bibinfo {author} {\bibfnamefont {A.}~\bibnamefont {Vishwanath}},\ }\bibfield  {title} {\bibinfo {title} {Riemannian geometry of resonant optical responses},\ }\href@noop {} {\bibfield  {journal} {\bibinfo  {journal} {Nature Physics}\ }\textbf {\bibinfo {volume} {18}},\ \bibinfo {pages} {290} (\bibinfo {year} {2022})}\BibitemShut {NoStop}%
\bibitem [{\citenamefont {Bouhon}\ \emph {et~al.}(2023)\citenamefont {Bouhon}, \citenamefont {Timmel},\ and\ \citenamefont {Slager}}]{Bouhon2023}%
  \BibitemOpen
  \bibfield  {author} {\bibinfo {author} {\bibfnamefont {A.}~\bibnamefont {Bouhon}}, \bibinfo {author} {\bibfnamefont {A.}~\bibnamefont {Timmel}},\ and\ \bibinfo {author} {\bibfnamefont {R.-J.}\ \bibnamefont {Slager}},\ }\href@noop {} {\bibinfo {title} {Quantum geometry beyond projective single bands}} (\bibinfo {year} {2023}),\ \Eprint {https://arxiv.org/abs/2303.02180} {arXiv:2303.02180 [cond-mat.mes-hall]} \BibitemShut {NoStop}%
\bibitem [{\citenamefont {Holder}\ \emph {et~al.}(2020)\citenamefont {Holder}, \citenamefont {Kaplan},\ and\ \citenamefont {Yan}}]{Holder2020}%
  \BibitemOpen
  \bibfield  {author} {\bibinfo {author} {\bibfnamefont {T.}~\bibnamefont {Holder}}, \bibinfo {author} {\bibfnamefont {D.}~\bibnamefont {Kaplan}},\ and\ \bibinfo {author} {\bibfnamefont {B.}~\bibnamefont {Yan}},\ }\bibfield  {title} {\bibinfo {title} {Consequences of time-reversal-symmetry breaking in the light-matter interaction: Berry curvature, quantum metric, and diabatic motion},\ }\href {https://doi.org/10.1103/PhysRevResearch.2.033100} {\bibfield  {journal} {\bibinfo  {journal} {Phys. Rev. Res.}\ }\textbf {\bibinfo {volume} {2}},\ \bibinfo {pages} {033100} (\bibinfo {year} {2020})}\BibitemShut {NoStop}%
\bibitem [{\citenamefont {Kaplan}\ \emph {et~al.}(2023)\citenamefont {Kaplan}, \citenamefont {Holder},\ and\ \citenamefont {Yan}}]{Kaplan2023}%
  \BibitemOpen
  \bibfield  {author} {\bibinfo {author} {\bibfnamefont {D.}~\bibnamefont {Kaplan}}, \bibinfo {author} {\bibfnamefont {T.}~\bibnamefont {Holder}},\ and\ \bibinfo {author} {\bibfnamefont {B.}~\bibnamefont {Yan}},\ }\bibfield  {title} {\bibinfo {title} {{Unifying semiclassics and quantum perturbation theory at nonlinear order}},\ }\href {https://doi.org/10.21468/SciPostPhys.14.4.082} {\bibfield  {journal} {\bibinfo  {journal} {SciPost Phys.}\ }\textbf {\bibinfo {volume} {14}},\ \bibinfo {pages} {082} (\bibinfo {year} {2023})}\BibitemShut {NoStop}%
\bibitem [{\citenamefont {Jankowski}\ and\ \citenamefont {Slager}(2024)}]{Wojciech2024}%
  \BibitemOpen
  \bibfield  {author} {\bibinfo {author} {\bibfnamefont {W.~J.}\ \bibnamefont {Jankowski}}\ and\ \bibinfo {author} {\bibfnamefont {R.-J.}\ \bibnamefont {Slager}},\ }\href@noop {} {\bibinfo {title} {Quantized shift response in multi-gap topological phases}} (\bibinfo {year} {2024}),\ \Eprint {https://arxiv.org/abs/2402.13245} {arXiv:2402.13245 [cond-mat.mes-hall]} \BibitemShut {NoStop}%
\bibitem [{\citenamefont {Kruchkov}\ and\ \citenamefont {Ryu}(2023)}]{Kruchkov2023}%
  \BibitemOpen
  \bibfield  {author} {\bibinfo {author} {\bibfnamefont {A.}~\bibnamefont {Kruchkov}}\ and\ \bibinfo {author} {\bibfnamefont {S.}~\bibnamefont {Ryu}},\ }\href@noop {} {\bibinfo {title} {Spectral sum rules reflect topological and quantum-geometric invariants}} (\bibinfo {year} {2023}),\ \Eprint {https://arxiv.org/abs/2312.17318} {arXiv:2312.17318 [cond-mat.str-el]} \BibitemShut {NoStop}%
\bibitem [{\citenamefont {Onishi}\ and\ \citenamefont {Fu}(2024{\natexlab{a}})}]{onishi2024quantum}%
  \BibitemOpen
  \bibfield  {author} {\bibinfo {author} {\bibfnamefont {Y.}~\bibnamefont {Onishi}}\ and\ \bibinfo {author} {\bibfnamefont {L.}~\bibnamefont {Fu}},\ }\bibfield  {title} {\bibinfo {title} {Quantum weight},\ }\href@noop {} {\bibfield  {journal} {\bibinfo  {journal} {arXiv preprint arXiv:2406.06783}\ } (\bibinfo {year} {2024}{\natexlab{a}})}\BibitemShut {NoStop}%
\bibitem [{\citenamefont {Onishi}\ and\ \citenamefont {Fu}(2024{\natexlab{b}})}]{onishi2024universal}%
  \BibitemOpen
  \bibfield  {author} {\bibinfo {author} {\bibfnamefont {Y.}~\bibnamefont {Onishi}}\ and\ \bibinfo {author} {\bibfnamefont {L.}~\bibnamefont {Fu}},\ }\bibfield  {title} {\bibinfo {title} {Universal relation between energy gap and dielectric constant},\ }\href@noop {} {\bibfield  {journal} {\bibinfo  {journal} {arXiv preprint arXiv:2401.04180}\ } (\bibinfo {year} {2024}{\natexlab{b}})}\BibitemShut {NoStop}%
\bibitem [{\citenamefont {Onishi}\ and\ \citenamefont {Fu}(2024{\natexlab{c}})}]{onishi2024fundamental}%
  \BibitemOpen
  \bibfield  {author} {\bibinfo {author} {\bibfnamefont {Y.}~\bibnamefont {Onishi}}\ and\ \bibinfo {author} {\bibfnamefont {L.}~\bibnamefont {Fu}},\ }\bibfield  {title} {\bibinfo {title} {Fundamental bound on topological gap},\ }\href@noop {} {\bibfield  {journal} {\bibinfo  {journal} {Physical Review X}\ }\textbf {\bibinfo {volume} {14}},\ \bibinfo {pages} {011052} (\bibinfo {year} {2024}{\natexlab{c}})}\BibitemShut {NoStop}%
\bibitem [{\citenamefont {Bradlyn}\ and\ \citenamefont {Abbamonte}(2024)}]{Bradlyn2024}%
  \BibitemOpen
  \bibfield  {author} {\bibinfo {author} {\bibfnamefont {B.}~\bibnamefont {Bradlyn}}\ and\ \bibinfo {author} {\bibfnamefont {P.}~\bibnamefont {Abbamonte}},\ }\href {https://arxiv.org/abs/2404.16144} {\bibinfo {title} {Spectral density and sum rules for second-order response functions}} (\bibinfo {year} {2024}),\ \Eprint {https://arxiv.org/abs/2404.16144} {arXiv:2404.16144 [cond-mat.mes-hall]} \BibitemShut {NoStop}%
\bibitem [{\citenamefont {Bernevig}(2013)}]{Bernevig2013}%
  \BibitemOpen
  \bibfield  {author} {\bibinfo {author} {\bibfnamefont {B.~A.}\ \bibnamefont {Bernevig}},\ }\href {https://doi.org/doi:10.1515/9781400846733} {\emph {\bibinfo {title} {{Topological Insulators and Topological Superconductors}}}}\ (\bibinfo  {publisher} {Princeton University Press},\ \bibinfo {address} {Princeton},\ \bibinfo {year} {2013})\BibitemShut {NoStop}%
\bibitem [{\citenamefont {Romero}\ \emph {et~al.}(2024)\citenamefont {Romero}, \citenamefont {Velasquez},\ and\ \citenamefont {Vergara}}]{romero2024n}%
  \BibitemOpen
  \bibfield  {author} {\bibinfo {author} {\bibfnamefont {J.}~\bibnamefont {Romero}}, \bibinfo {author} {\bibfnamefont {C.~A.}\ \bibnamefont {Velasquez}},\ and\ \bibinfo {author} {\bibfnamefont {J.~D.}\ \bibnamefont {Vergara}},\ }\bibfield  {title} {\bibinfo {title} {$ n $-bein formalism for the parameter space of quantum geometry},\ }\href@noop {} {\bibfield  {journal} {\bibinfo  {journal} {arXiv preprint arXiv:2406.19468}\ } (\bibinfo {year} {2024})}\BibitemShut {NoStop}%
\bibitem [{\citenamefont {Roy}(2014)}]{Roy2014}%
  \BibitemOpen
  \bibfield  {author} {\bibinfo {author} {\bibfnamefont {R.}~\bibnamefont {Roy}},\ }\bibfield  {title} {\bibinfo {title} {Band geometry of fractional topological insulators},\ }\href {https://doi.org/10.1103/PhysRevB.90.165139} {\bibfield  {journal} {\bibinfo  {journal} {Phys. Rev. B}\ }\textbf {\bibinfo {volume} {90}},\ \bibinfo {pages} {165139} (\bibinfo {year} {2014})}\BibitemShut {NoStop}%
\bibitem [{\citenamefont {Ledwith}\ \emph {et~al.}(2023)\citenamefont {Ledwith}, \citenamefont {Vishwanath},\ and\ \citenamefont {Parker}}]{Ledwith2023}%
  \BibitemOpen
  \bibfield  {author} {\bibinfo {author} {\bibfnamefont {P.~J.}\ \bibnamefont {Ledwith}}, \bibinfo {author} {\bibfnamefont {A.}~\bibnamefont {Vishwanath}},\ and\ \bibinfo {author} {\bibfnamefont {D.~E.}\ \bibnamefont {Parker}},\ }\bibfield  {title} {\bibinfo {title} {{Vortexability: A unifying criterion for ideal fractional Chern insulators}},\ }\href {https://doi.org/10.1103/PhysRevB.108.205144} {\bibfield  {journal} {\bibinfo  {journal} {Phys. Rev. B}\ }\textbf {\bibinfo {volume} {108}},\ \bibinfo {pages} {205144} (\bibinfo {year} {2023})}\BibitemShut {NoStop}%
\bibitem [{\citenamefont {Peotta}\ and\ \citenamefont {T\"orm\"a}(2015)}]{Peotta2015}%
  \BibitemOpen
  \bibfield  {author} {\bibinfo {author} {\bibfnamefont {S.}~\bibnamefont {Peotta}}\ and\ \bibinfo {author} {\bibfnamefont {P.}~\bibnamefont {T\"orm\"a}},\ }\bibfield  {title} {\bibinfo {title} {{Superfluidity in Topologically Nontrivial Flat Bands}},\ }\href {https://doi.org/10.1038/ncomms9944} {\bibfield  {journal} {\bibinfo  {journal} {Nat. Commun.}\ }\textbf {\bibinfo {volume} {6}},\ \bibinfo {pages} {8944} (\bibinfo {year} {2015})}\BibitemShut {NoStop}%
\bibitem [{\citenamefont {Wang}\ \emph {et~al.}(2023)\citenamefont {Wang}, \citenamefont {Kaplan}, \citenamefont {Zhang}, \citenamefont {Holder}, \citenamefont {Cao}, \citenamefont {Wang}, \citenamefont {Zhou}, \citenamefont {Zhou}, \citenamefont {Jiang}, \citenamefont {Zhang}, \citenamefont {Ru}, \citenamefont {Cai}, \citenamefont {Watanabe}, \citenamefont {Taniguchi}, \citenamefont {Yan},\ and\ \citenamefont {Gao}}]{Wang2023}%
  \BibitemOpen
  \bibfield  {author} {\bibinfo {author} {\bibfnamefont {N.}~\bibnamefont {Wang}}, \bibinfo {author} {\bibfnamefont {D.}~\bibnamefont {Kaplan}}, \bibinfo {author} {\bibfnamefont {Z.}~\bibnamefont {Zhang}}, \bibinfo {author} {\bibfnamefont {T.}~\bibnamefont {Holder}}, \bibinfo {author} {\bibfnamefont {N.}~\bibnamefont {Cao}}, \bibinfo {author} {\bibfnamefont {A.}~\bibnamefont {Wang}}, \bibinfo {author} {\bibfnamefont {X.}~\bibnamefont {Zhou}}, \bibinfo {author} {\bibfnamefont {F.}~\bibnamefont {Zhou}}, \bibinfo {author} {\bibfnamefont {Z.}~\bibnamefont {Jiang}}, \bibinfo {author} {\bibfnamefont {C.}~\bibnamefont {Zhang}}, \bibinfo {author} {\bibfnamefont {S.}~\bibnamefont {Ru}}, \bibinfo {author} {\bibfnamefont {H.}~\bibnamefont {Cai}}, \bibinfo {author} {\bibfnamefont {K.}~\bibnamefont {Watanabe}}, \bibinfo {author} {\bibfnamefont {T.}~\bibnamefont {Taniguchi}}, \bibinfo {author} {\bibfnamefont {B.}~\bibnamefont {Yan}},\ and\ \bibinfo {author} {\bibfnamefont {W.}~\bibnamefont {Gao}},\ }\bibfield  {title}
  {\bibinfo {title} {{Quantum-metric-induced nonlinear transport in a topological antiferromagnet}},\ }\href {https://doi.org/10.1038/s41586-023-06363-3} {\bibfield  {journal} {\bibinfo  {journal} {Nature}\ }\textbf {\bibinfo {volume} {621}},\ \bibinfo {pages} {487} (\bibinfo {year} {2023})}\BibitemShut {NoStop}%
\bibitem [{\citenamefont {Sodemann}\ and\ \citenamefont {Fu}(2015)}]{Sodemann2015}%
  \BibitemOpen
  \bibfield  {author} {\bibinfo {author} {\bibfnamefont {I.}~\bibnamefont {Sodemann}}\ and\ \bibinfo {author} {\bibfnamefont {L.}~\bibnamefont {Fu}},\ }\bibfield  {title} {\bibinfo {title} {{Quantum Nonlinear Hall Effect Induced by Berry Curvature Dipole in Time-Reversal Invariant Materials}},\ }\href {https://doi.org/10.1103/PhysRevLett.115.216806} {\bibfield  {journal} {\bibinfo  {journal} {Phys. Rev. Lett.}\ }\textbf {\bibinfo {volume} {115}},\ \bibinfo {pages} {216806} (\bibinfo {year} {2015})}\BibitemShut {NoStop}%
\bibitem [{\citenamefont {Het{\'e}nyi}\ and\ \citenamefont {L{\'e}vay}(2023)}]{hetenyi2023fluctuations}%
  \BibitemOpen
  \bibfield  {author} {\bibinfo {author} {\bibfnamefont {B.}~\bibnamefont {Het{\'e}nyi}}\ and\ \bibinfo {author} {\bibfnamefont {P.}~\bibnamefont {L{\'e}vay}},\ }\bibfield  {title} {\bibinfo {title} {Fluctuations, uncertainty relations, and the geometry of quantum state manifolds},\ }\href@noop {} {\bibfield  {journal} {\bibinfo  {journal} {Physical Review A}\ }\textbf {\bibinfo {volume} {108}},\ \bibinfo {pages} {032218} (\bibinfo {year} {2023})}\BibitemShut {NoStop}%
\bibitem [{\citenamefont {Onishi}\ \emph {et~al.}(2024)\citenamefont {Onishi}, \citenamefont {Avdoshkin},\ and\ \citenamefont {Fu}}]{onishi2024geometric}%
  \BibitemOpen
  \bibfield  {author} {\bibinfo {author} {\bibfnamefont {Y.}~\bibnamefont {Onishi}}, \bibinfo {author} {\bibfnamefont {A.}~\bibnamefont {Avdoshkin}},\ and\ \bibinfo {author} {\bibfnamefont {L.}~\bibnamefont {Fu}},\ }\bibfield  {title} {\bibinfo {title} {Geometric bound on structure factor},\ }\href@noop {} {\bibfield  {journal} {\bibinfo  {journal} {arXiv preprint arXiv:2412.02656}\ } (\bibinfo {year} {2024})}\BibitemShut {NoStop}%
\bibitem [{\citenamefont {Jankowski}\ \emph {et~al.}(2024)\citenamefont {Jankowski}, \citenamefont {Morris}, \citenamefont {Davoyan}, \citenamefont {Bouhon}, \citenamefont {{\"U}nal},\ and\ \citenamefont {Slager}}]{jankowski2024non}%
  \BibitemOpen
  \bibfield  {author} {\bibinfo {author} {\bibfnamefont {W.~J.}\ \bibnamefont {Jankowski}}, \bibinfo {author} {\bibfnamefont {A.~S.}\ \bibnamefont {Morris}}, \bibinfo {author} {\bibfnamefont {Z.}~\bibnamefont {Davoyan}}, \bibinfo {author} {\bibfnamefont {A.}~\bibnamefont {Bouhon}}, \bibinfo {author} {\bibfnamefont {F.~N.}\ \bibnamefont {{\"U}nal}},\ and\ \bibinfo {author} {\bibfnamefont {R.-J.}\ \bibnamefont {Slager}},\ }\bibfield  {title} {\bibinfo {title} {{Non-Abelian Hopf-Euler insulators}},\ }\href@noop {} {\bibfield  {journal} {\bibinfo  {journal} {Physical Review B}\ }\textbf {\bibinfo {volume} {110}},\ \bibinfo {pages} {075135} (\bibinfo {year} {2024})}\BibitemShut {NoStop}%
\bibitem [{\citenamefont {Shi}\ \emph {et~al.}(2021)\citenamefont {Shi}, \citenamefont {Zhang}, \citenamefont {Chang},\ and\ \citenamefont {Song}}]{Shi2021}%
  \BibitemOpen
  \bibfield  {author} {\bibinfo {author} {\bibfnamefont {L.-k.}\ \bibnamefont {Shi}}, \bibinfo {author} {\bibfnamefont {D.}~\bibnamefont {Zhang}}, \bibinfo {author} {\bibfnamefont {K.}~\bibnamefont {Chang}},\ and\ \bibinfo {author} {\bibfnamefont {J.~C.~W.}\ \bibnamefont {Song}},\ }\bibfield  {title} {\bibinfo {title} {{Geometric Photon-Drag Effect and Nonlinear Shift Current in Centrosymmetric Crystals}},\ }\href {https://doi.org/10.1103/PhysRevLett.126.197402} {\bibfield  {journal} {\bibinfo  {journal} {Phys. Rev. Lett.}\ }\textbf {\bibinfo {volume} {126}},\ \bibinfo {pages} {197402} (\bibinfo {year} {2021})}\BibitemShut {NoStop}%
\bibitem [{\citenamefont {Wang}\ \emph {et~al.}(2022)\citenamefont {Wang}, \citenamefont {Tang}, \citenamefont {Li},\ and\ \citenamefont {Qian}}]{Wang2022}%
  \BibitemOpen
  \bibfield  {author} {\bibinfo {author} {\bibfnamefont {H.}~\bibnamefont {Wang}}, \bibinfo {author} {\bibfnamefont {X.}~\bibnamefont {Tang}}, \bibinfo {author} {\bibfnamefont {J.}~\bibnamefont {Li}},\ and\ \bibinfo {author} {\bibfnamefont {X.}~\bibnamefont {Qian}},\ }\bibfield  {title} {\bibinfo {title} {{Generalized Wilson loop method for nonlinear light-matter interaction}},\ }\href {https://doi.org/10.1038/s41535-022-00472-4} {\bibfield  {journal} {\bibinfo  {journal} {npj Qantum Materials}\ }\textbf {\bibinfo {volume} {7}},\ \bibinfo {pages} {61} (\bibinfo {year} {2022})}\BibitemShut {NoStop}%
\bibitem [{\citenamefont {Zhu}\ and\ \citenamefont {Alexandradinata}(2024)}]{Zhu2024}%
  \BibitemOpen
  \bibfield  {author} {\bibinfo {author} {\bibfnamefont {P.}~\bibnamefont {Zhu}}\ and\ \bibinfo {author} {\bibfnamefont {A.}~\bibnamefont {Alexandradinata}},\ }\bibfield  {title} {\bibinfo {title} {Anomalous shift and optical vorticity in the steady photovoltaic current},\ }\href {https://doi.org/10.1103/PhysRevB.110.115108} {\bibfield  {journal} {\bibinfo  {journal} {Phys. Rev. B}\ }\textbf {\bibinfo {volume} {110}},\ \bibinfo {pages} {115108} (\bibinfo {year} {2024})}\BibitemShut {NoStop}%
\bibitem [{\citenamefont {Sipe}\ and\ \citenamefont {Shkrebtii}(2000)}]{Sipe2000}%
  \BibitemOpen
  \bibfield  {author} {\bibinfo {author} {\bibfnamefont {J.~E.}\ \bibnamefont {Sipe}}\ and\ \bibinfo {author} {\bibfnamefont {A.~I.}\ \bibnamefont {Shkrebtii}},\ }\bibfield  {title} {\bibinfo {title} {Second-order optical response in semiconductors},\ }\href {https://doi.org/10.1103/PhysRevB.61.5337} {\bibfield  {journal} {\bibinfo  {journal} {Phys. Rev. B}\ }\textbf {\bibinfo {volume} {61}},\ \bibinfo {pages} {5337} (\bibinfo {year} {2000})}\BibitemShut {NoStop}%
\bibitem [{\citenamefont {Mitscherling}\ and\ \citenamefont {Metzner}(2018)}]{Mitscherling2018}%
  \BibitemOpen
  \bibfield  {author} {\bibinfo {author} {\bibfnamefont {J.}~\bibnamefont {Mitscherling}}\ and\ \bibinfo {author} {\bibfnamefont {W.}~\bibnamefont {Metzner}},\ }\bibfield  {title} {\bibinfo {title} {{Longitudinal conductivity and Hall coefficient in two-dimensional metals with spiral magnetic order}},\ }\bibfield  {journal} {\bibinfo  {journal} {Physical Review B}\ }\textbf {\bibinfo {volume} {98}},\ \href {https://doi.org/10.1103/physrevb.98.195126} {10.1103/physrevb.98.195126} (\bibinfo {year} {2018})\BibitemShut {NoStop}%
\bibitem [{\citenamefont {King-Smith}\ and\ \citenamefont {Vanderbilt}(1993)}]{Kingsmith1993}%
  \BibitemOpen
  \bibfield  {author} {\bibinfo {author} {\bibfnamefont {R.~D.}\ \bibnamefont {King-Smith}}\ and\ \bibinfo {author} {\bibfnamefont {D.}~\bibnamefont {Vanderbilt}},\ }\bibfield  {title} {\bibinfo {title} {Theory of polarization of crystalline solids},\ }\href@noop {} {\bibfield  {journal} {\bibinfo  {journal} {Phys. Rev. B}\ }\textbf {\bibinfo {volume} {47}},\ \bibinfo {pages} {1651} (\bibinfo {year} {1993})}\BibitemShut {NoStop}%
\bibitem [{\citenamefont {Resta}\ and\ \citenamefont {Vanderbilt}(2007)}]{Resta2007}%
  \BibitemOpen
  \bibfield  {author} {\bibinfo {author} {\bibfnamefont {R.}~\bibnamefont {Resta}}\ and\ \bibinfo {author} {\bibfnamefont {D.}~\bibnamefont {Vanderbilt}},\ }\bibinfo {title} {Theory of polarization: A modern approach},\ in\ \href {https://doi.org/10.1007/978-3-540-34591-6_2} {\emph {\bibinfo {booktitle} {Physics of Ferroelectrics: A Modern Perspective}}}\ (\bibinfo  {publisher} {Springer Berlin Heidelberg},\ \bibinfo {address} {Berlin, Heidelberg},\ \bibinfo {year} {2007})\ pp.\ \bibinfo {pages} {31--68}\BibitemShut {NoStop}%
\bibitem [{\citenamefont {Patankar}\ \emph {et~al.}(2018)\citenamefont {Patankar}, \citenamefont {Wu}, \citenamefont {Lu}, \citenamefont {Rai}, \citenamefont {Tran}, \citenamefont {Morimoto}, \citenamefont {Parker}, \citenamefont {Grushin}, \citenamefont {Nair}, \citenamefont {Analytis}, \citenamefont {Moore}, \citenamefont {Orenstein},\ and\ \citenamefont {Torchinsky}}]{Patankar2018}%
  \BibitemOpen
  \bibfield  {author} {\bibinfo {author} {\bibfnamefont {S.}~\bibnamefont {Patankar}}, \bibinfo {author} {\bibfnamefont {L.}~\bibnamefont {Wu}}, \bibinfo {author} {\bibfnamefont {B.}~\bibnamefont {Lu}}, \bibinfo {author} {\bibfnamefont {M.}~\bibnamefont {Rai}}, \bibinfo {author} {\bibfnamefont {J.~D.}\ \bibnamefont {Tran}}, \bibinfo {author} {\bibfnamefont {T.}~\bibnamefont {Morimoto}}, \bibinfo {author} {\bibfnamefont {D.~E.}\ \bibnamefont {Parker}}, \bibinfo {author} {\bibfnamefont {A.~G.}\ \bibnamefont {Grushin}}, \bibinfo {author} {\bibfnamefont {N.~L.}\ \bibnamefont {Nair}}, \bibinfo {author} {\bibfnamefont {J.~G.}\ \bibnamefont {Analytis}}, \bibinfo {author} {\bibfnamefont {J.~E.}\ \bibnamefont {Moore}}, \bibinfo {author} {\bibfnamefont {J.}~\bibnamefont {Orenstein}},\ and\ \bibinfo {author} {\bibfnamefont {D.~H.}\ \bibnamefont {Torchinsky}},\ }\bibfield  {title} {\bibinfo {title} {Resonance-enhanced optical nonlinearity in the weyl semimetal taas},\ }\href {https://doi.org/10.1103/PhysRevB.98.165113}
  {\bibfield  {journal} {\bibinfo  {journal} {Phys. Rev. B}\ }\textbf {\bibinfo {volume} {98}},\ \bibinfo {pages} {165113} (\bibinfo {year} {2018})}\BibitemShut {NoStop}%
\bibitem [{\citenamefont {Jiang}\ \emph {et~al.}(2025)\citenamefont {Jiang}, \citenamefont {Holder},\ and\ \citenamefont {Yan}}]{Jiang2025}%
  \BibitemOpen
  \bibfield  {author} {\bibinfo {author} {\bibfnamefont {Y.}~\bibnamefont {Jiang}}, \bibinfo {author} {\bibfnamefont {T.}~\bibnamefont {Holder}},\ and\ \bibinfo {author} {\bibfnamefont {B.}~\bibnamefont {Yan}},\ }\href {https://arxiv.org/abs/2503.04943} {\bibinfo {title} {{Revealing Quantum Geometry in Nonlinear Quantum Materials}}} (\bibinfo {year} {2025}),\ \Eprint {https://arxiv.org/abs/2503.04943} {arXiv:2503.04943 [cond-mat.mes-hall]} \BibitemShut {NoStop}%
\bibitem [{\citenamefont {Gao}\ \emph {et~al.}(2021)\citenamefont {Gao}, \citenamefont {Addison}, \citenamefont {Mele},\ and\ \citenamefont {Rappe}}]{Gao2021}%
  \BibitemOpen
  \bibfield  {author} {\bibinfo {author} {\bibfnamefont {L.}~\bibnamefont {Gao}}, \bibinfo {author} {\bibfnamefont {Z.}~\bibnamefont {Addison}}, \bibinfo {author} {\bibfnamefont {E.~J.}\ \bibnamefont {Mele}},\ and\ \bibinfo {author} {\bibfnamefont {A.~M.}\ \bibnamefont {Rappe}},\ }\bibfield  {title} {\bibinfo {title} {Intrinsic fermi-surface contribution to the bulk photovoltaic effect},\ }\href {https://doi.org/10.1103/PhysRevResearch.3.L042032} {\bibfield  {journal} {\bibinfo  {journal} {Phys. Rev. Res.}\ }\textbf {\bibinfo {volume} {3}},\ \bibinfo {pages} {L042032} (\bibinfo {year} {2021})}\BibitemShut {NoStop}%
\end{thebibliography}%

\end{document}